\documentclass[final, twocolumn]{article}

\usepackage[tt=false,type1=true]{libertine}
\usepackage[varqu]{zi4}
\usepackage[libertine]{newtxmath}
\usepackage[T1]{fontenc}
\usepackage[utf8]{inputenc}
\usepackage{amsmath}
\usepackage{amssymb}
\usepackage{mathtools}
\usepackage{microtype}
\usepackage[inline]{enumitem}
\usepackage{multirow}
\usepackage{booktabs}
\usepackage{subcaption}
\usepackage[usenames,dvipsnames,table]{xcolor}
\usepackage{nag}
\usepackage{todonotes}
\usepackage{hyperref}
\usepackage{glossaries-extra}

\setabbreviationstyle[acronym]{long-short}
\glssetcategoryattribute{acronym}{nohyperfirst}{true}

\usepackage{placeins}
\usepackage{pdfrender}
\usepackage{varwidth}
\usepackage[margin=3cm]{geometry}

\newacronym{PBR}{PBR}{Physically Based Rendering}
\newacronym{LTE}{LTE}{Light transport equation} 
\newacronym{NEE}{NEE}{Next Event Estimation}
\newacronym{SSS}{SSS}{Subsurface scattering}
\newacronym{PT}{PT}{Path tracer}
\newacronym{VOLPATH_SIMPLE}{VOLPATH SIMPLE}{Simple volumetric path tracer}
\newacronym{VOLPATH}{VOLPATH}{Extended volumetric path tracer}
\newacronym{BDPT}{BDPT}{Bidirectional path tracer}
\newacronym{PM}{PM}{Photon mapper}
\newacronym{PPM}{PPM}{Progressive photon mapper}
\newacronym{SPPM}{SPPM}{Stochastic progressive photon mapper}
\newacronym{PSSMLT}{PSSMLT}{Primary Sample Space Metropolis Light Transport}
\newacronym{MLT}{MLT}{Metropolis Light Transport, Path Space Metropolis Light Transport}
\newacronym{ERPT}{ERPT}{Energy Redistribution Path tracing}
\newacronym{IRRCACHE}{IRRCACHE}{Irradiance caching}
\newacronym{spp}{spp}{Samples per pixel}
\newacronym{BRDF}{BRDF}{Bidirectional reflectance distribution function}

\newacronym{mc}{MC}{Monte Carlo}
\newacronym{odpr}{ODPR}{Online Database for Physically-based Rendering}
	
\newcommand*{\boldcheckmark}{%
	\textpdfrender{
		TextRenderingMode=FillStroke,
		LineWidth=.8pt,
	}{\checkmark}%
}

\title{Test Scene Design for Physically Based Rendering}
\author{Elias Brugger, Christian Freude, Michael Wimmer}
\date{TU Wien}

\begin{document}

\maketitle

\begin{abstract}
Physically based rendering is a discipline in computer graphics which aims at reproducing certain light and material appearances that occur in the real world. Complex scenes can be difficult to compute for rendering algorithms. This paper introduces a new comprehensive test database of scenes that treat different light setups in conjunction with diverse materials and discusses its design principles. A lot of research is focused on the development of new algorithms that can deal with difficult light conditions and materials efficiently. This database delivers a comprehensive foundation for evaluating existing and newly developed rendering techniques. A final evaluation compares different results of different rendering algorithms for all scenes. The full data set can be downloaded from \textit{Zenodo}\footnote{\url{https://zenodo.org/}, 10.5281/zenodo.4002021} and individual scenes can be accessed via a dedicated \textit{web repository}\footnote{\url{https://odpr.cg.tuwien.ac.at}}.
\end{abstract}

\section{Introduction}

\subsection{Motivation}

\emph{\gls{PBR}} is a discipline in computer graphics that aims at producing images that resemble the real world as accurate as possible. \gls{PBR} is integral to photorealism in movies and games as well as architectural and industrial visualization.

In general, the techniques of \gls{PBR} aim to simulate the lighting conditions as they appear in reality. Therefore, rendering such images requires simulation of light, which is often computationally expensive and time consuming. Generally, \gls{PBR} consists of several techniques that have to play together to achieve a realistic image, most importantly \emph{path tracing} and associated \emph{Monte Carlo Integration} are fundamental methods for creating such images. \cite{PharrEtAl2016}

Different \emph{Monte Carlo}-based algorithms exist to render physically correct images, each of them having advantages and drawbacks depending on the scene setup. Research work is focused to improve the efficiency of \emph{Monte Carlo Rendering}. Newly developed approaches need to be compared and evaluated against already existing ones. The problem is that there are many different scattered scenes available but there is no commonly used database for evaluation and testing, this makes it difficult to compare methods.

The goal of this paper is to deliver a step towards a more consistently designed scene database for the evaluation of different aspects of \gls{PBR} and \emph{Monte Carlo Rendering}. Since there can be considerable differences in efficiency and capabilities between the different algorithms, the scenes are designed to include a variety of challenging cases that try to enhance and show the characteristics of such algorithms.

This paper will discuss the individual test scenes and the underlying design principles
and also includes rendered examples using a selection of different algorithms.
Additional sections regarding \gls{mc} rendering are included in the appendix.
The next section will give an overview of existing test scene databases and related work.

\subsection{Related Work}

The database described in this paper lays a comprehensive foundation for testing light and material phenomena. For further testing with more complex models, material, and light setups, refer to the \gls{PBR} database created and gathered by Morgan McGuire \cite{McGuire2017Data}. It contains several complex scenes with large amounts of polygons as well as a variety of different materials and light setups.

Another comprehensive and more complex database was created and gathered by Benedikt Bitterli \cite{BitterliScenes}. This database contains 32 different scenes ranging from simple scenes to complex scenes with complex light setup and 3D objects. Partially overlapping scenes can be found in the \gls{PBR} database by Wenzel et al. \cite{PBRTScenes}. This dataset features extensive representations of outdoor vegetation with a variety of unique plant models as well as scenes with large portions of glass materials. 

An initial approach on scene creation for \emph{Global Illumination} can be found in the paper \emph{Global Illumination Test Scenes} by Smits et al. \cite{SmitsTestScenes}. The scenes contain elementary 3D objects and monochrome light setups to show basic phenomena of light transport. 

An analytical approach on \emph{Global Illumination} scenes can be found in the paper \emph{Testing Monte-Carlo global illumination methods with analytically computable scenes} by Szirmay-Kalos et al. \cite{LaszloScenes}, which investigates the correctness of \emph{Global Illumination} algorithms. Because \emph{Monte Carlo Rendering} uses random sampling, correct or incorrect results (because of implementation errors) are often not noticeable. As a solution, scenes with known exact solutions are used to verify the algorithms. 

Another approach on verification of \gls{PBR} algorithms can be found in the paper \emph{Verification of Physically Based Rendering Algorithms} by Ulbricht et al. \cite{ulbricht-2006-VOP}. This paper discusses several state of the art approaches to verify the correctness of light transport simulation as well as advantages and disadvantages of them. 

The database discussed in this paper, in contrast, should give a more structured and comprehensive set of test scenes used to observe a variety of phenomena in \gls{PBR} without focus on physical or analytical correctness. Moreover, the modular library containing the 3D objects delivers the possibility to adjust the existing scenes or even create completely new ones.

\section{Scene Database}
\subsection{Introduction}
In this chapter, the main motivation of this paper, the creation of a scene database, will be discussed and presented in detail. Twenty scenes were created out of a library of different objects including different lighting situations, simple objects, and more complex objects with different material setups. Since not all algorithms achieve the same result with the same render time, it is interesting to observe the different lighting and material conditions in order to choose an appropriate algorithm that delivers a good result and does not take too long in terms of render time. The evaluation part of this paper will compare render results of different \emph{integrator}s (algorithms like \emph{bidirectional path tracing }or \emph{Metropolis light transport}) in terms of render time and visual quality.

The render engine that is used to render the scenes is \emph{mitsuba}, developed by Jakob Wenzel \cite{Mitsuba2010} in 2010. As this is the most common render system for scientific purposes, it is an appropriate choice to observe realistic materials and light conditions and show difficulties and issues common in the field of \gls{PBR}.

\emph{mitsuba} is a comprehensive renderer that combines multiple techniques from important research work in \emph{Physically Based Rendering}. Many inspirations in the system come from the book \cite{PharrEtAl2016} that was introduced earlier. Important \emph{integrator}s and material properties will be explained later. The material presets are based on real-world measurements according to the \emph{mitsuba} documentation \cite{MitsubaDoc2014}. The documentation is an extensive collection of all the materials and light emitters that come with the render system, furthermore it contains the usage and constraints of the different components. The scenes have been modeled in Blender 2.79 with the \emph{mitsuba} plugin. Conveniently, the scenes can be exported as XML file, as the native \emph{mitsuba} application operates with this file system.

\subsubsection{Conceptual Approach}

There are several criteria that accompanied the creation of the test scenes. The first and most important one was to design them as representative and expressive as possible. That means that the advantages and drawbacks of the different algorithms can be well distinguished based on the renderings. More specifically, the idea was to use different advanced materials like glass and rough surfaces in order to increase the complexity of the calculations and to challenge the algorithms. 

Moreover, the scenes feature typical phenomena/effects characteristic to \gls{PBR}. However, certain scenes are designed to particularly concentrate on few effects. Refractions and reflections play an important role in realistic synthesized images. Translucent and reflective materials are not perfectly smooth in reality, that means that they have certain surface irregularities. Several scenes feature such surfaces. The difference is that rough materials generally demand more complex calculations, because light scatters more in the scene.

More sophisticated scenes feature materials with \gls{SSS}, which can only be rendered with a few of the available algorithms. \gls{SSS} materials are considerably relevant to \gls{PBR}, because -- in the real world -- many objects are partially translucent. In regard to \emph{participating media}, some scenes feature atmospheric-like particles, which is a challenge for any algorithm but is also integral to realistic renderings.

Several scenes also feature complex geometry in conjunction with refractive and reflective materials. This will generally challenge the calculations of light paths, especially when it comes to resolving caustics. That is where certain \emph{integrator}s have an advantage compared to basic algorithms like \emph{path tracing}. Table \ref{tab:scenes} shows a list with the scenes and the corresponding effects they focus on.

\begin{table*}[h]
	\centering
	\begin{tabular}{cccccccccc}
		\toprule
		Scene & RefrSpec & RefrGl & Reflec & Caust & SoftSh & ColBl & PartM & Cgeom & SSS \\
		\cmidrule(lr){1-1}
		\cmidrule{2-10}
		\ref{section:water_glass} & $\boldcheckmark$ & & \checkmark & $\boldcheckmark$ & $\boldcheckmark$ & & & \checkmark \\
		\cmidrule(lr){1-1}
		\cmidrule{2-10}
		\ref{section:glass_pendulum} & $\boldcheckmark$ & & \checkmark & $\boldcheckmark$ & \checkmark  \\
		\cmidrule(lr){1-1}
		\cmidrule{2-10}
		\ref{section:glass_pendulum_different_shapes} & $\boldcheckmark$ & & \checkmark & $\boldcheckmark$ & \checkmark & & $\boldcheckmark$ \\
		\cmidrule(lr){1-1}
		\cmidrule{2-10}
		\ref{section:sphere_and_lenses} & $\boldcheckmark$ & & \checkmark & $\boldcheckmark$ & \checkmark  \\
		\cmidrule(lr){1-1}
		\cmidrule{2-10}
		\ref{section:water_caustics_1} & $\boldcheckmark$ & & \checkmark & $\boldcheckmark$ & \checkmark & $\boldcheckmark$ & & $\boldcheckmark$ \\
		\cmidrule(lr){1-1}
		\cmidrule{2-10}
		\ref{section:water_caustics_2} & $\boldcheckmark$ & & \checkmark & $\boldcheckmark$ & \checkmark & \checkmark & $\boldcheckmark$ & $\boldcheckmark$ \\
		\cmidrule(lr){1-1}
		\cmidrule{2-10}
		\ref{section:color_bleeding} & & & $\boldcheckmark$ & & $\boldcheckmark$ & $\boldcheckmark$ \\
		\cmidrule(lr){1-1}
		\cmidrule{2-10}
		\ref{section:smooth_and_rough_glass} & $\boldcheckmark$ & $\boldcheckmark$ &\checkmark & $\boldcheckmark$ & \checkmark & & & $\boldcheckmark$ \\
		\cmidrule(lr){1-1}
		\cmidrule{2-10}
		\ref{section:sss_complex} & \checkmark & & \checkmark & & \checkmark & & & $\boldcheckmark$ & $\boldcheckmark$ \\
		\cmidrule(lr){1-1}
		\cmidrule{2-10}
		\ref{section:sss_complex_and_water} & $\boldcheckmark$ & & \checkmark & \checkmark & \checkmark & & & $\boldcheckmark$ & $\boldcheckmark$  \\
		\cmidrule(lr){1-1}
		\cmidrule{2-10}
		\ref{section:lens_light_transmission} & $\boldcheckmark$ & & $\boldcheckmark$ & $\boldcheckmark$ & \checkmark & & $\boldcheckmark$ & $\boldcheckmark$  \\
		\cmidrule(lr){1-1}
		\cmidrule{2-10}
		\ref{section:the_lens_effect} & $\boldcheckmark$ & & \checkmark & & & \checkmark & & \checkmark  \\
		\cmidrule(lr){1-1}
		\cmidrule{2-10}
		\ref{section:double_mirrors} & \checkmark & & $\boldcheckmark$ & \checkmark & \checkmark & & & $\boldcheckmark$  \\
		\cmidrule(lr){1-1}
		\cmidrule{2-10}
		\ref{section:glass_prism} & $\boldcheckmark$ & & $\boldcheckmark$ & $\boldcheckmark$ & \checkmark  \\
		\cmidrule(lr){1-1}
		\cmidrule{2-10}
		\ref{section:shadows_of_different_light_sources} & & & \checkmark & & $\boldcheckmark$ & & & \checkmark \\
		\cmidrule(lr){1-1}
		\cmidrule{2-10}
		\ref{section:light_from_slightly_opened_door} & $\boldcheckmark$ & & $\boldcheckmark$ & & $\boldcheckmark$  \\
		\cmidrule(lr){1-1}
		\cmidrule{2-10}
		\ref{section:light_from_outside} & $\boldcheckmark$ & & $\boldcheckmark$ & & $\boldcheckmark$ & & & $\boldcheckmark$  \\
		\cmidrule(lr){1-1}
		\cmidrule{2-10}
		\ref{section:sss_and_realistic_light_setup} & $\boldcheckmark$ & $\boldcheckmark$ & \checkmark & & & & & $\boldcheckmark$ & $\boldcheckmark$ \\
		\cmidrule(lr){1-1}
		\cmidrule{2-10}
		\ref{section:complex_sss_and_rough_glass_surfaces} & $\boldcheckmark$  & $\boldcheckmark$  & $\boldcheckmark$  & $\boldcheckmark$ & \checkmark & \checkmark & & $\boldcheckmark$  & $\boldcheckmark$   \\
		\cmidrule(lr){1-1}
		\cmidrule{2-10}
		\ref{section:water_caustics_and_broad_light_distribution} & $\boldcheckmark$ & & \checkmark & $\boldcheckmark$ & \checkmark & \checkmark & & $\boldcheckmark$ & $\boldcheckmark$  \\
		\bottomrule
	\end{tabular}
	\caption{An overview of the different aspects covered by each scene. Bold check marks indicate primary effects and normal check marks indicate secondary effects which are not in the main focus of the respective scenes. RefrSpec -- Refraction specular, RefrGl -- Refraction glossy, Reflec -- Reflection, Caust -- Caustics, SoftSh -- Soft shadows, ColBl -- Color bleeding, PartM -- \emph{Participating media}, Cgeom -- Complex geometry, \gls{SSS} -- Subsurface scattering}
	\label{tab:scenes} %
\end{table*}

\subsection{Scene -- Water glass}\label{section:water_glass}

\begin{figure}[!htbp]
	\centering
	\includegraphics[width=\linewidth]{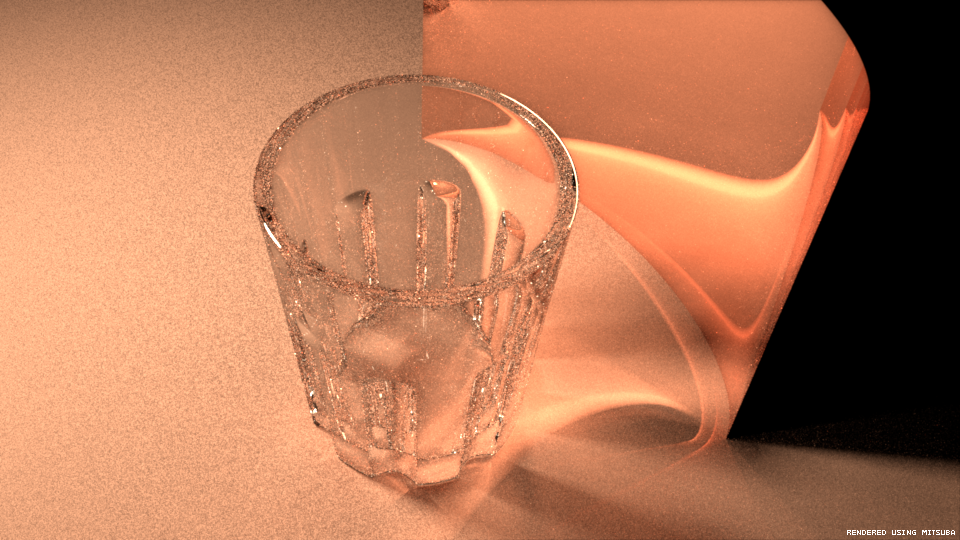}
	\caption{Simple scene with a water glass and a mirror causing different caustics and reflections. Rendered with \gls{PSSMLT} and \emph{Independent sampler} with 2048 samples per pixel (\gls{spp}).}
	\label{fig:caustic_direct_light} %
\end{figure}

Figure \ref{fig:caustic_direct_light} shows a simple scene with a water glass and a bent mirror. It was made primarily to show complex caustics of the glass material and the mirror. The glass has a common glass material and several indents on the side. The mirror to the right is bent and applied with a copper material preset. The ground has a common metallic material with a high roughness coefficient (see Section \ref{subsubsection:rough_conductor_material}). The scene involves a sphere light in the top left of the scene. The sphere light itself is a spherical shape with a material defined as area emitter (see Section \ref{subsubsection:area_emitter}), this is a typical setup for any kind of realistic light source in \emph{mitsuba}.

The glass itself is relatively hard to render because the light that penetrates the glass causes many reflections in several different directions. In order to simulate the real world as accurate as possible, the 3D objects have to have several properties. The glass is defined as an object with an inner material. \emph{mitsuba} materials allow to set properties for materials that are in the exterior and in the interior (exterior is the side of the normal directions) of the outer boundary. In case of the glass, the exterior is defined as air and the interior as a glass with a typical index of refraction, while the glass material itself is a \emph{Smooth dielectric material} \cite{MitsubaDoc2014} (see Section \ref{subsubsection:smooth_dielectric_material}). The renderer needs to know these parameters to compute the correct reflection angles, and depending on the preset, the materials have different indices of refraction. 

The ground of the scene appears relatively grainy, that is because the surface material has a high roughness coefficient, which means a widespread light scattering in the scene. The scene was primarily designed to test different \emph{integrator}s and to evaluate which of them can handle caustics well. It appears that the caustics coming from the mirror are less grainier than those coming from the glass.

\subsubsection{Area Emitter} \label{subsubsection:area_emitter}
An area emitter is a light source that emits light via the exterior surface of an arbitrary shape. Unlike some other lights definable in the \emph{mitsuba} renderer, area emitters generally cast soft shadows. \cite{MitsubaDoc2014}

\subsubsection{Smooth dielectric material}
\label{subsubsection:smooth_dielectric_material}
\begin{figure}[!htbp]
	\centering
	\includegraphics[width=0.25\textwidth]{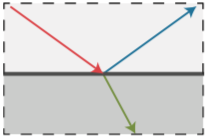}
	\caption{The red arrow is the incident light ray, the blue arrow depicts the reflected ray and the green arrow shows the refracted light through the glass material. \cite{MitsubaDoc2014}}
	\label{fig:smooth_dielectric_material} %
\end{figure}

Figure \ref{fig:smooth_dielectric_material} shows a geometric interpretation of the \emph{Smooth dielectric material} in \emph{mitsuba}. It defines a material with a perfectly smooth surface, that means that the BSDF scatters light in discrete directions (as opposed to rough surfaces with a continuum of resulting light rays). The green arrow in Figure \ref{fig:smooth_dielectric_material} shows the change of direction of the incoming light (red arrow) depending on the index of refraction. The blue arrow shows the portion of the light that is reflected on the surface.

\subsection{Scene -- Glass pendulum}\label{section:glass_pendulum}
\begin{figure}[!htbp]
	\centering
	\includegraphics[width=\linewidth]{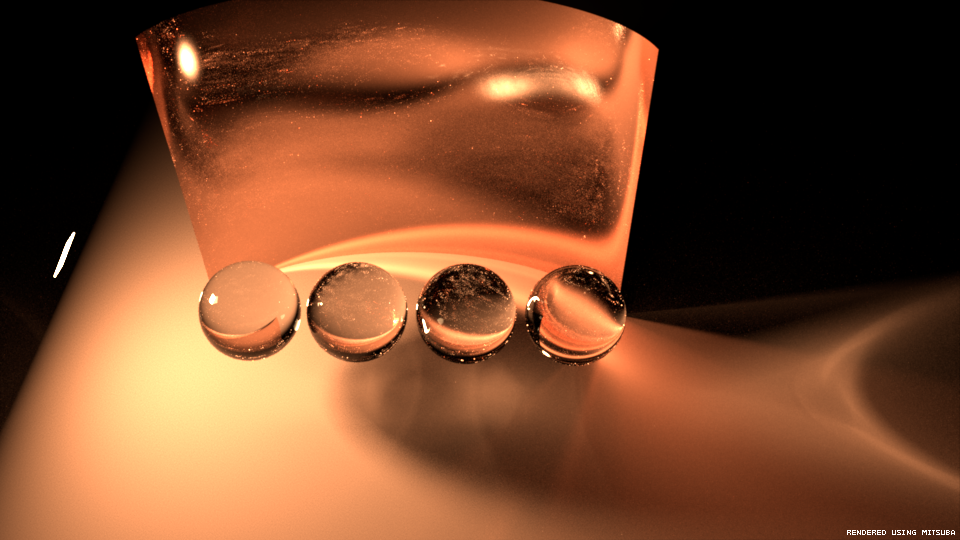}
	\caption{Glass spheres arranged as pendulum to show light propagation and caustics. Rendered with \gls{PSSMLT} and \emph{Independent sampler} with 1024 \gls{spp}.}
	\label{fig:caustic_direct_light_pendulum} %
\end{figure}

Figure \ref{fig:caustic_direct_light_pendulum} shows a scene with four glass spheres with a crown glass material preset arranged in a row which is primarily designed to show light propagation through glass materials as well as caustics of spherical shapes. It simulates a light pendulum that transmits the incoming light from the first until the last sphere. An area light connected to a common plane serves as light source. It is slightly raised to have the light direction oriented towards the ground. As a result, the light not only shines straight through the spheres, it also causes the spheres to throw caustics onto the ground. The light propagation of spheres is shown in detail in Figure \ref{fig:sphere_caustics}. However, the arrangement of the spheres leads to light being bundled at most for the first sphere. The subsequent spheres gradually receive less light from the previous one. Interesting to observe are the caustics that appear on the ground. They come from the spheres but do not have a typical spherical shape as it is common for such shapes. This comes from the variation of light directly falling on the ground and the intersection with subsequent spheres. This leads to light being refracted in another direction. 

\begin{figure}[!htbp]
	\centering
	\includegraphics[width=0.25\textwidth]{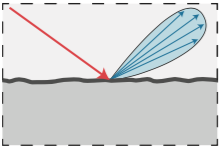}
	\caption{Light (red arrow) falling onto a surface with tiny fluctuations results in a reflected continuum of scattered light (blue array). \cite{Mitsuba2010}}
	\label{fig:rought_conductor_material}
\end{figure}

A glossy bent mirror is placed behind the pendulum. It has a copper material with little roughness. This is achieved through a \emph{Rough conductor material} that, unlike the \emph{Smooth conductor material}, has surface scattering as seen in Figure \ref{fig:rought_conductor_material}. Interestingly, the ground itself is not as grainy as the ground in Figure \ref{fig:caustic_direct_light} although the surface has the same material with equal roughness. This may come from the complexity of the glass portions in the scene and from the shape and intensity of the light source.

\subsubsection{Rough conductor material}
\label{subsubsection:rough_conductor_material}

The \emph{Rough conductor material} describes a microfacet distribution of a surface. The implementation in \emph{mitsuba} is based on the paper \emph{Microfacet Models for Refraction through Rough Surfaces} by Walter et al. \cite{Walt2007}. Microfacets describe surfaces with micro geometry. For instance a metal surface in the real world is not perfectly smooth but it has tiny surface fluctuations. The microfacet model attempts to simulate such surfaces through varying surface normals. The light computation for rough surfaces generally takes longer than for perfectly smooth materials \cite{MitsubaDoc2014}.

\subsection{Scene -- Glass pendulum different shapes}\label{section:glass_pendulum_different_shapes}
\begin{figure}[!htbp]
	\centering
	\includegraphics[width=\linewidth]{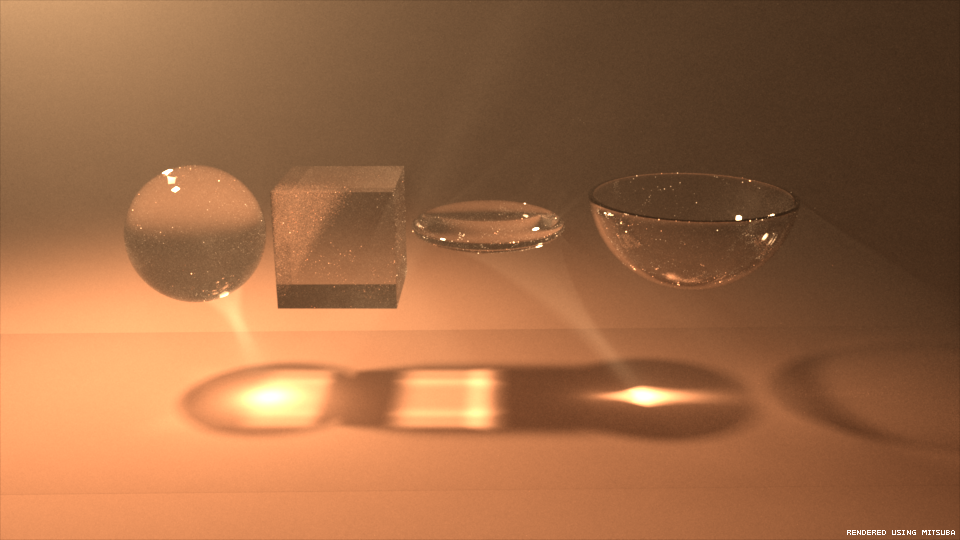}
	\caption{Different glass shapes arranged causing different caustic shapes.  Rendered with \gls{PSSMLT} and \emph{Independent sampler} with 1024 \gls{spp}.}
	\label{fig:caustic_direct_light_pendulum_pm}
\end{figure}

\begin{figure}[!htbp]
	\centering
	\includegraphics[width=0.5\textwidth]{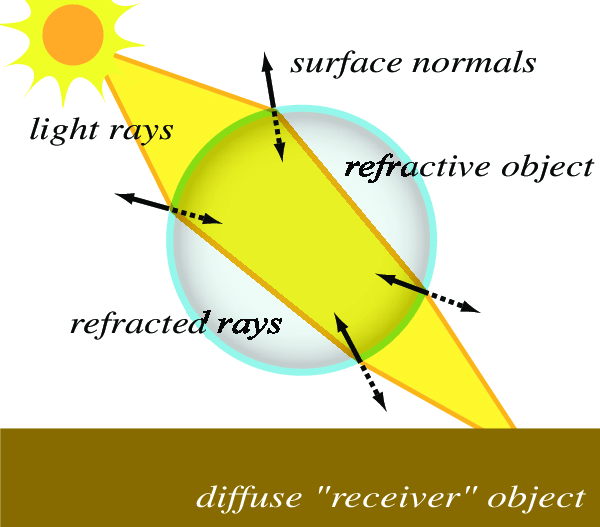}
	\caption{Light propagation through a translucent sphere. \cite{Shah2007}}
	\label{fig:sphere_caustics} %
\end{figure}

Figure \ref{fig:caustic_direct_light_pendulum_pm} shows a similar scene like before. This time, different shapes are arranged in a row. The light source is a plane emitting from the top left of the arrangement. As different glass shapes throw different caustics, this is an appropriate setup to compare some common shapes. In order to give the light rays an appearance, a \emph{participating medium} was put around the arrangement. The bounding box of the medium can be seen on the ground where the hard edges appear. The inner surface appears a bit darker because of the light being gradually absorbed/reflected by the particles in the medium. Participating media is good for showing how light rays are cast through space. This can be observed well by looking at the light cones that emerge from underneath the sphere and the glass lens in the middle. The glass cube and the glass bowl have caustics as well but the bundling of the light rays is too weak to appear in the medium, but they can be observed on the ground. The size difference of the light cone of the sphere and the lens is remarkable. This is because of the different diameters. This phenomenon can be seen in Figure \ref{fig:sphere_caustics}.

\subsection{Scene -- Sphere and lenses}\label{section:sphere_and_lenses}
\begin{figure}[!htbp]
	\centering
	\includegraphics[width=\linewidth]{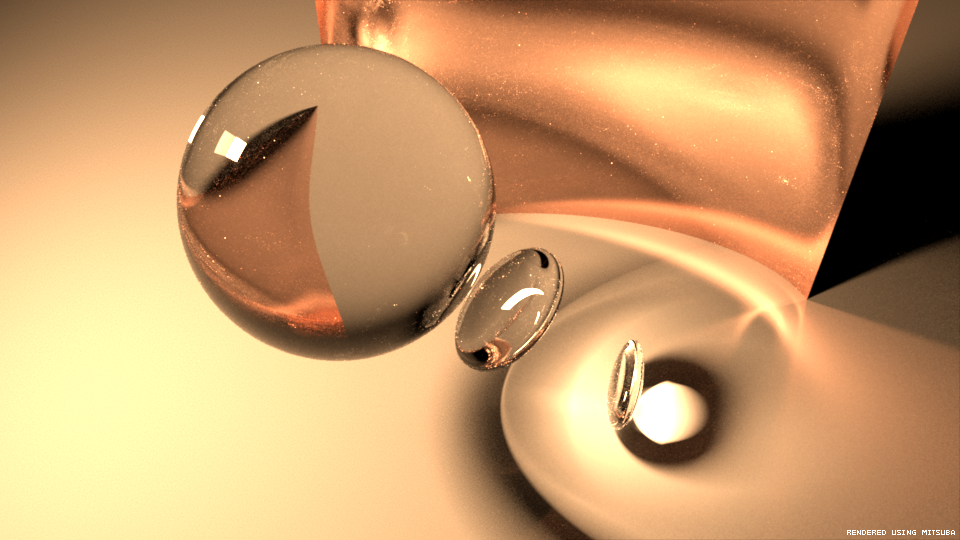}
	\caption{Glass sphere and glass lenses forwarding light rays differently.  Rendered with \gls{PSSMLT} and \emph{Independent sampler} with 1024 \gls{spp}.}
	\label{fig:caustic_direct_light_sphere} %
\end{figure}

The scene in Figure \ref{fig:caustic_direct_light_sphere} consists of a larger glass sphere and two tinier lenses. The main focus in this scene is to show the properties of light bundling and what shapes are designed to achieve it. The background again features a bent mirror with a \emph{Rough conductor material}. The light source comes from the top left of the scene. The main focus here is on the light rays that are propagated through the glass materials. 

The glass sphere transmits a relatively large amount of the incoming light to the first lens. The lens then casts the incoming portions of light onto the last lens. It is noticeable that the light intensity differs considerably. When looking at the ground to the left of the rendering, it can be observed that due to the transmission and refraction of the light rays through the glass materials, the light gradually intensifies. On the right side of the last lens on the ground, the light intensity is at its peak.

\subsection{Scene -- Water caustics 1}\label{section:water_caustics_1}
\begin{figure}[!htbp]
	\centering
	\includegraphics[width=\linewidth]{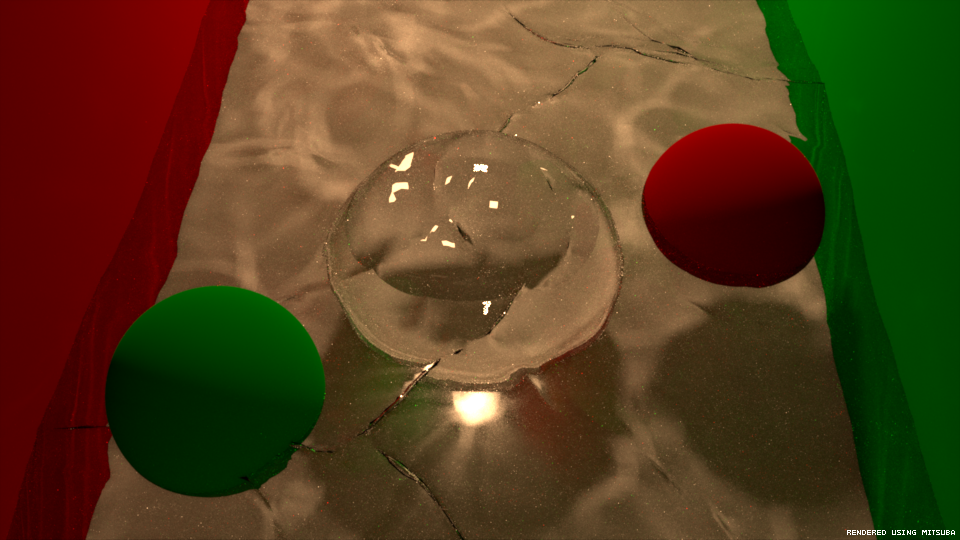}
	\caption{Glass sphere in water with waves causing different complex caustics.  Rendered with \gls{PSSMLT} and \emph{Independent sampler} with 1024 \gls{spp}.}
	\label{fig:caustic_direct_light_waterwaves} %
\end{figure}

The scene in Figure \ref{fig:caustic_direct_light_waterwaves} was designed to show caustics of waves in a medium like water. An area light is placed on the top and behind the objects. Furthermore, a glass sphere is placed in the middle as well as two diffuse spheres on the sides. 

The colored spheres and the walls on the side of the rendering have a \emph{Smooth diffuse material} \cite{MitsubaDoc2014}. It can be described as a perfectly diffuse material with even distribution of light in any direction (see Section \ref{subsubsection:smooth_diffuse_material}). This material leads to light reflections that are independent from the point of view \cite{MitsubaDoc2014}.

Specifically hard to render are the caustics coming from the waves of the water. The colored objects in the scene should help visualize the distribution of the reflected light. Color bleeding resulting from the spheres can be seen underneath the glass sphere. In the real world, there are generally no perfectly smooth diffuse materials, but this was set up out of simplicity.

\subsubsection{Smooth diffuse material}
\label{subsubsection:smooth_diffuse_material}
\begin{figure}[!htbp]
	\centering
	\includegraphics[width=0.25\textwidth]{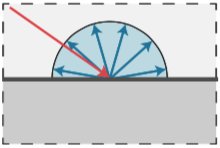}
	\caption{Light distribution of \emph{Smooth diffuse material}. The red arrow is the incident light and the blue continuum is the scattered light on the surface. \cite{MitsubaDoc2014}}
	\label{fig:smooth_diffuse_material} %
\end{figure}

The \emph{Smooth diffuse material} is a surface material that reflects and scatters light independently from the point of view, as seen in Figure \ref{fig:smooth_diffuse_material}.

\subsection{Scene -- Water caustics 2}\label{section:water_caustics_2}

\begin{figure}[!htbp]
	\centering
	\includegraphics[width=\linewidth]{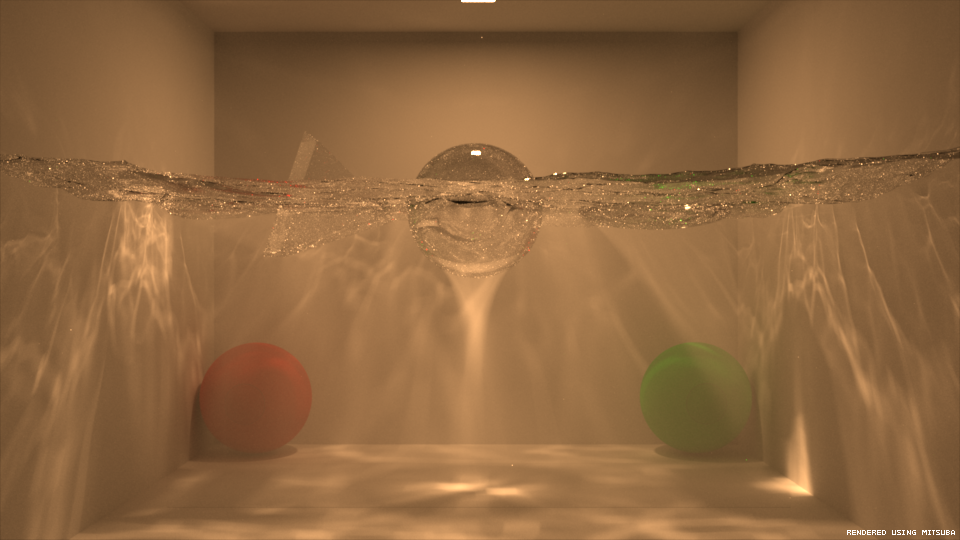}
	\caption{Room filled with water with three different glass shapes transmitting light into a partially participating medium. Rendered with \gls{PSSMLT} and \emph{Independent sampler} with 1024 \gls{spp}.}
	\label{fig:caustic_direct_light_waterwaves_pm} %
\end{figure}

The scene in Figure \ref{fig:caustic_direct_light_waterwaves_pm} shows similar effects like before. A room with a rough diffuse material is filled with water. The water is a deformed plane that simulates waves. Furthermore, a common glass sphere is placed in the middle. On the left there is also a glass prism as well as a glass lens on the right side. 

In this case, a \emph{participating medium} was added for the portions underneath the water surface to show the light rays that are thrown onto the ground. The two colored diffuse spheres in the corners in the background appear less saturated because of the caustics and the \emph{participating medium}.

Already known are the typical caustics of the sphere in the middle as well as the caustics of the lens. The prism has a more uncommon behavior in respect of light refraction and reflection, regardless, the light rays that are refracted are almost not visible in the medium. This is similar to the scene in Figure \ref{fig:caustic_direct_light_pendulum_pm}, where the caustics of the box and the bowl are not visible either, apart from the light that hits the ground.

The portions of the water throwing caustics are larger and more detailed. It is noticeable that the further away the appearances of the caustics are from the light source, the more blurred they become. This can be seen when comparing the caustics directly underneath the surface and the appearances on the ground. Naturally this depends on the edginess of the waves. Relatively choppy waves result in short focal points whereas broad waves result in longer focal points. This is the same principle as it can be observed when comparing the caustics of a lens and a sphere.

\subsection{Scene -- Color bleeding}\label{section:color_bleeding}
\begin{figure}[!htbp]
	\centering
	\includegraphics[width=\linewidth]{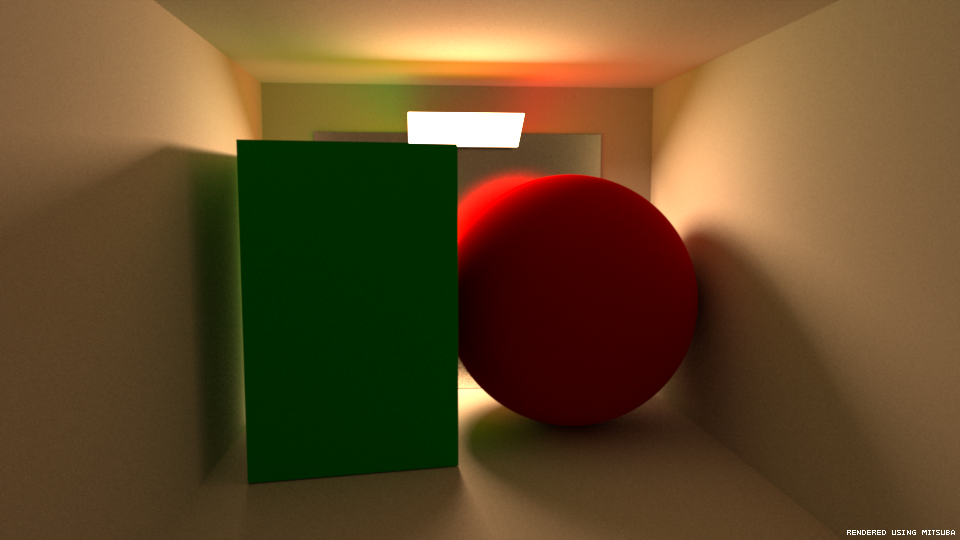}
	\caption{Scene showing emphasized color bleeding effect. Rendered with \gls{PSSMLT} and \emph{Independent sampler} with 1024 \gls{spp}.}
	\label{fig:colorbleeding} %
\end{figure}

The scene in Figure \ref{fig:colorbleeding} shows a scene set up to show color bleeding, which is a common effect in realistic computer graphics because of the exchange of indirect light between objects in the scene. The idea behind is based on the scene of the \emph{cornell boxes} from the original paper \emph{Modeling the Interaction of Light between Diffuse Surfaces} by Goral et al. \cite{GoralEtAl1984}.

A modification to the original rendering was made in terms of object shapes and the assignment of colors. In this case a green box and a red sphere are placed in front of a mirror on the back wall. The plane area light throws light onto the back side of the box and sphere. This causes the reflections to mainly throw colored reflections onto the back side of the room, which can be seen as green and red reflections around the light source. The mirror on the back wall has little roughness applied so the scattering of the light supports the visualization of color bleeding.

\subsection{Scene -- Smooth and rough glass}\label{section:smooth_and_rough_glass}
\begin{figure}[!htbp]
	\centering
	\includegraphics[width=\linewidth]{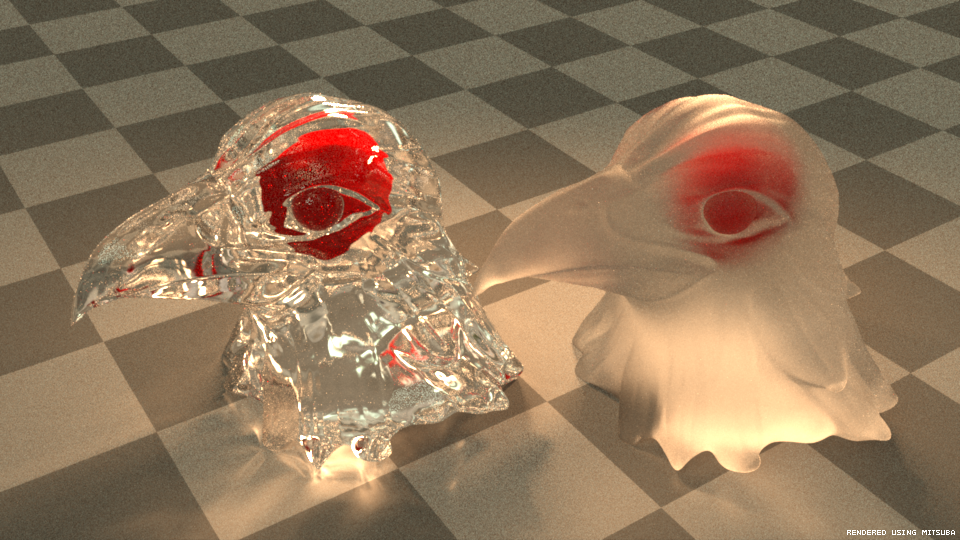}
	\caption{Two high polygon objects with glass material, the left one with no roughness and the right one with moderate roughness. Rendered with \gls{PSSMLT} and \emph{Independent sampler} with 1024 \gls{spp}.}
	\label{fig:complex_glass} %
\end{figure}

In the scene shown in Figure \ref{fig:complex_glass}, a high polygon object was used to simulate two different glass materials. The left version has a common glass material with no roughness at all (\emph{Smooth dielectric material}), whereas the right one has a \emph{Rough dielectric material} \cite{MitsubaDoc2014} (see Section \ref{subsubsection:rough_dielectric_material}). The complex object itself has around half a million polygons. A red diffuse sphere was placed inside each of the objects to emphasize the impacts of roughness. On top of each object there is an area light that casts light through the glass.

While the light rays for the \emph{Smooth dielectric material} reflect in discrete directions, the light rays for the \emph{Rough dielectric material} scatter in a continuum. On the one hand, the light gets scattered on the exterior of the surface, on the other hand, the light that passes the surface also results in a continuum of scattered light. In case of the rendering, this results in caustics on the ground being more visible for the left object. Additionally, the light that is scattered within the right object is much more present than for the left one because the roughness makes it collect more light inside the surface, specifically visible on the ground.

\subsubsection{Rough dielectric material}
\label{subsubsection:rough_dielectric_material}
\begin{figure}[!htbp]
	\centering
	\includegraphics[width=0.25\textwidth]{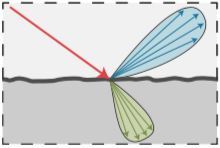}
	\caption{Light distribution of \emph{Rough dielectric material}. The incident light (red arrow) gets reflected in a continuum (blue array) as well as refracted in a continuum through the surface (green array). \cite{MitsubaDoc2014}}
	\label{fig:rough_dielectric_material} %
\end{figure}

As well as the rough version of the conductor material, the \emph{Rough dielectric material} (see Figure \ref{fig:rough_dielectric_material}) is also based on the theory from the paper \emph{Microfacet Models for Refraction through Rough Surfaces} \cite{Walt2007} and implements specific microfacet distributions for glass materials. \cite{MitsubaDoc2014}

\subsection{Scene -- \gls{SSS} complex}\label{section:sss_complex}
\begin{figure}[!htbp]
	\centering
	\includegraphics[width=\linewidth]{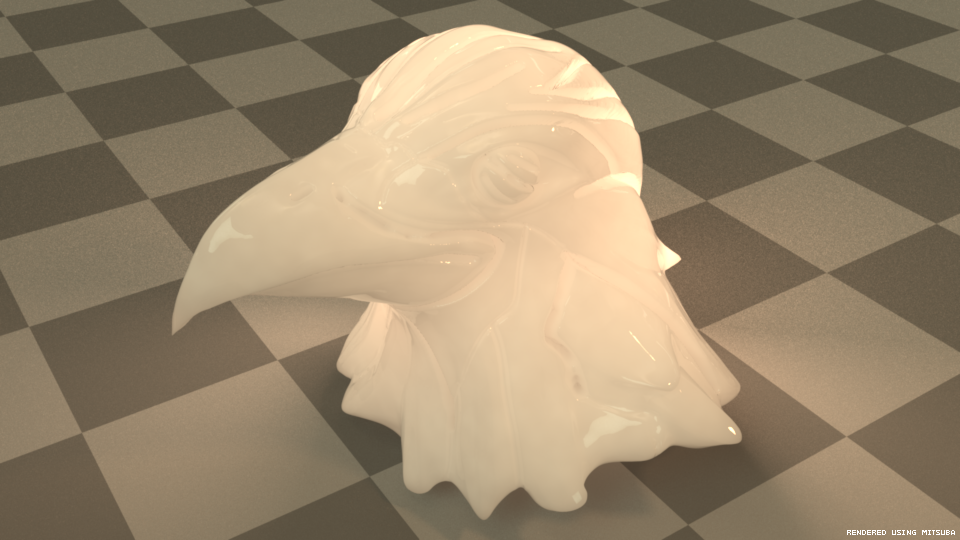}
	\caption{High polygon object with \emph{Dipole-based subsurface scattering} material. Rendered with \emph{Path tracer} and \emph{Independent sampler} with 1024 \gls{spp}.}
	\label{fig:complex_sss} %
\end{figure}

This scene shows again a complex object slightly modified by adding fissures all over it. The material is designed to show the properties of an \gls{SSS} material. In order to apply \gls{SSS} in \emph{mitsuba}, the base material has to be either a \emph{Smooth plastic material} or a \emph{Rough plastic material}. In case of the scene in Figure \ref{fig:complex_sss}, the material is a \emph{Smooth plastic material}, which is more complex than the materials used so far (see Section \ref{subsubsection:smooth_plastic_material}). With \gls{SSS}, materials like porcelain or wax can be simulated. Typically, light shining on such objects not only reflects it directly on the surface but it also lets portions of light through the surface. The inner medium of the object can be adjusted to have different scattering and absorption coefficients as well. This can be used to control how much light eventually shines through the object. It is clear that, the more solid an object is the less light will appear on the other side. To emphasize the effect of an \gls{SSS} material, an area light is placed behind the object. It can be seen that portions of the object are considerably bright because the thickness at this location is low.

Mentionable is that \gls{SSS} can be simulated in two ways in \emph{mitsuba}. The first one is an approximation (see Section \ref{subsubsection:dipole_sss}) of the medium that normally is in the interior of the object. The second option is to actually set a \emph{participating medium} in the interior of the object. The main advantage of the approximation method is a better performance but at costs of accuracy. \emph{Participating media} is the preferable option to simulate real \gls{SSS}, but it usually has a higher render time.

\subsubsection{Smooth plastic material}
\label{subsubsection:smooth_plastic_material}
\begin{figure}[!htbp]
	\centering
	\includegraphics[width=0.25\textwidth]{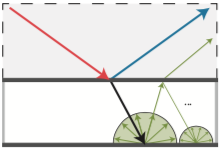}
	\caption{Light distribution of \emph{Smooth plastic material} The incident light (red arrow) results in a reflection on the surface (blue arrow) and several scattering events within the subsurface (black arrow and green arrays/arrow). \cite{MitsubaDoc2014}}
	\label{fig:smooth_plastic_material}
\end{figure}

The \emph{Smooth plastic material} simulates a surface with a subsurface, as seen in Figure \ref{fig:smooth_plastic_material}. It simulates a diffuse surface with a dielectric coating surrounding it. As a result, there are generally many internal scattering events. On the one hand, the light gets reflected (blue arrow) when it hits the dielectric coating, on the other hand, the remaining portion of the light (black arrow) goes through the surface and hits the diffuse base layer. This triggers the internal scattering process (green arrays and arrows within the subsurface). \cite{MitsubaDoc2014}

\subsubsection{Dipole-based subsurface scattering model}
\label{subsubsection:dipole_sss}
This \gls{SSS} model is implemented in \emph{mitsuba} and has its origins in radiative transport \cite{Eason2001} and medical physics \cite{Peng2012}. The implementation is based on the article \emph{A practical model for subsurface light transport} by Jensen et al. \cite{Jensen2001}. \cite{MitsubaDoc2014}

The \emph{Smooth plastic material} (see Section \ref{subsubsection:smooth_plastic_material}) in conjunction with the \emph{Dipole-based subsurface scattering model}, an approximate simulation of \gls{SSS} can be achieved. This is done by setting the diffuse reflectance parameter of the plastic material to zero and have the dipole plugin calculate the diffuse part instead. \cite{MitsubaDoc2014}

\subsection{Scene -- \gls{SSS} complex and water}\label{section:sss_complex_and_water}
\begin{figure}[!htbp]
	\centering
	\includegraphics[width=\linewidth]{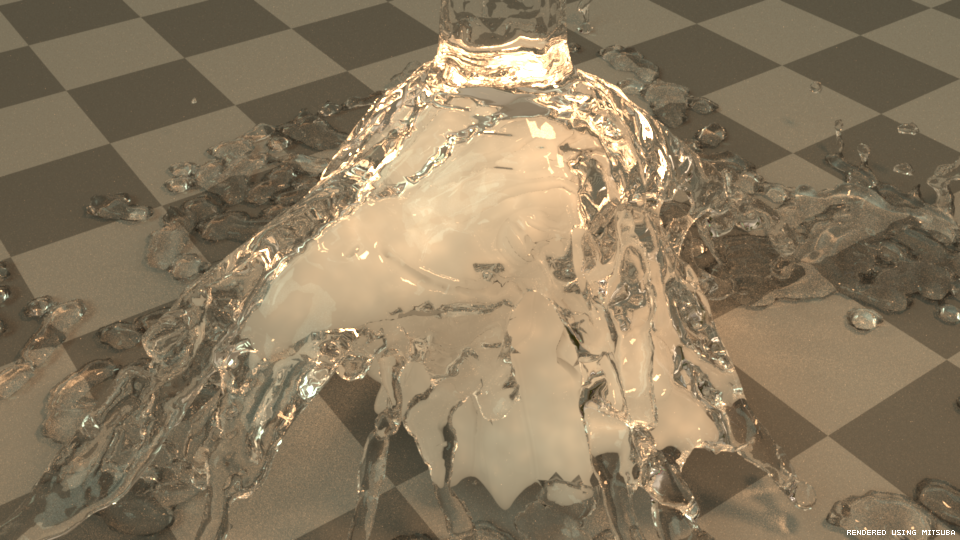}
	\caption{High polygon object with \emph{Dipole-based subsurface scattering} material with additional water outflow coming from above. Rendered with \emph{Path tracer} and \emph{Independent sampler} with 1024 \gls{spp}.}
	\label{fig:complex_raven_watercoat} %
\end{figure}

The scene in Figure \ref{fig:complex_raven_watercoat} is a modified version of the scene in Figure \ref{fig:complex_sss}. This time, a simulated water outflow has been placed above the complex object. The idea behind it is to observe how light behaves in combination with water and a subsurface material underneath. An area light is placed behind the complex object as well.

Since the scene contains only materials with zero roughness -- apart from the ground -- the light processing is relatively straight forward and simple to render. This is the reason why there is almost no grain in the rendering compared to other scenes. The water itself has a relatively low resolution in terms of polygon count to simplify the scene. Regardless, the water appears fairly realistic although the viscosity of the water seems to be higher, which is rather atypical. Also mind the little caustics in the bottom right corner coming from the water slipping off of the object surface. Because only a few \emph{integrator}s are able to render scenes containing subsurface materials, caustics are often difficult to render. This problem will be treated more detailed later on. 

As an example, this scene has been rendered with the common \emph{path tracer} \emph{integrator}. This \emph{integrator} usually has troubles rendering caustics and therefore \emph{integrator}s like \emph{bidirectional path tracer} are commonly used. The issue with the bidirectional method is that it cannot handle \gls{SSS} materials in \emph{mitsuba} \cite{MitsubaDoc2014}, as this is the case in this scene.

\subsection{Scene -- Lens light transmission}\label{section:lens_light_transmission}
\begin{figure}[!htbp]
	\centering
	\includegraphics[width=\linewidth]{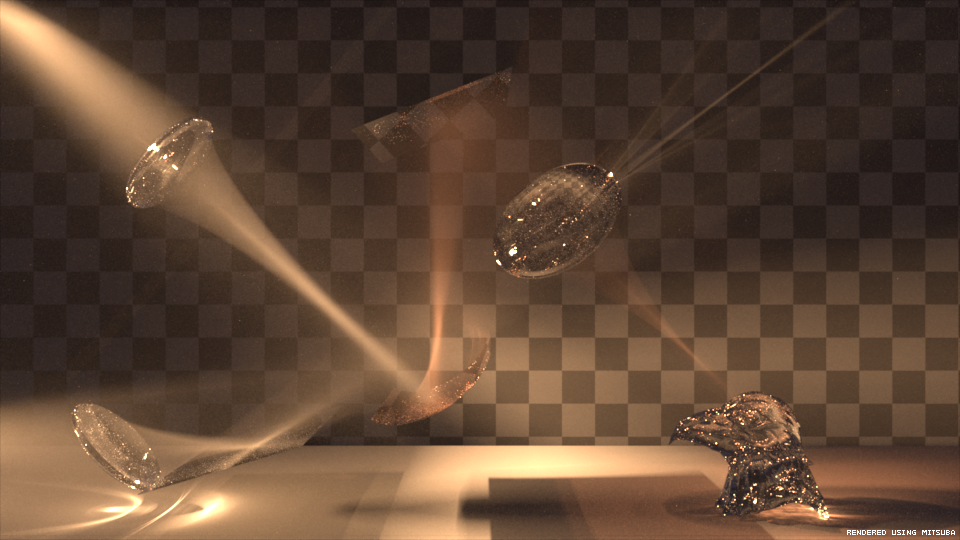}
	\caption{More complex scene showing the properties of lenses of difference thickness resulting in different ray propagations. Rendered with \gls{PSSMLT} and \emph{Independent sampler} with 1024 \gls{spp}.}
	\label{fig:lenses_convex} %
\end{figure}

The scene in Figure \ref{fig:lenses_convex} shows a setup designed to show the effects of applying a different thickness to convex lenses. Two light sources are placed in the top left corner and in the bottom left corner.

To catch as much light as possible, the light sources are surrounded by geometry, this makes it possible to control the size of the light cone. The first lens in the top left portion of the rendering receives almost all of the incoming light from the top light source. As the lens propagates the incoming light rays onto the bent mirror in about the middle part of the rendering, the rays once more get bundled to construct a relatively narrow focal point. The straight mirror at the mid top receives the scattered light from the bent mirror and reflects it towards a lens with a higher thickness. It is noticeable that the light reaching the last lens is already considerably low. Regardless, the lens bundles the light once more to throw a caustic with a slightly higher light intensity onto a high polygon object, which shows the effect of a convex lens with a higher thickness.

The lens in the bottom left corner and the corresponding mirror on the right side is built to transfer light towards the side of the last lens that is described above. This results in several tiny transmitted rays emerging from the top right of the lens and converging towards the top right of the rendering. Interesting to observe are the considerably long focal points of the lights rays transmitting the lens. The color from the light being reflected by the bent mirror is destined to show clearly what portions of the light get reflected to what location. In order to achieve light rays appearing in a scene, again a \emph{participating medium} was setup to facilitate this intent. Considerably interesting are the caustics resulting from the bottom left lens, since it appears to throw a variety of light rays from different directions.

\subsection{Scene -- The lens effect}\label{section:the_lens_effect}
\begin{figure}[!htbp]
	\centering
	\includegraphics[width=\linewidth]{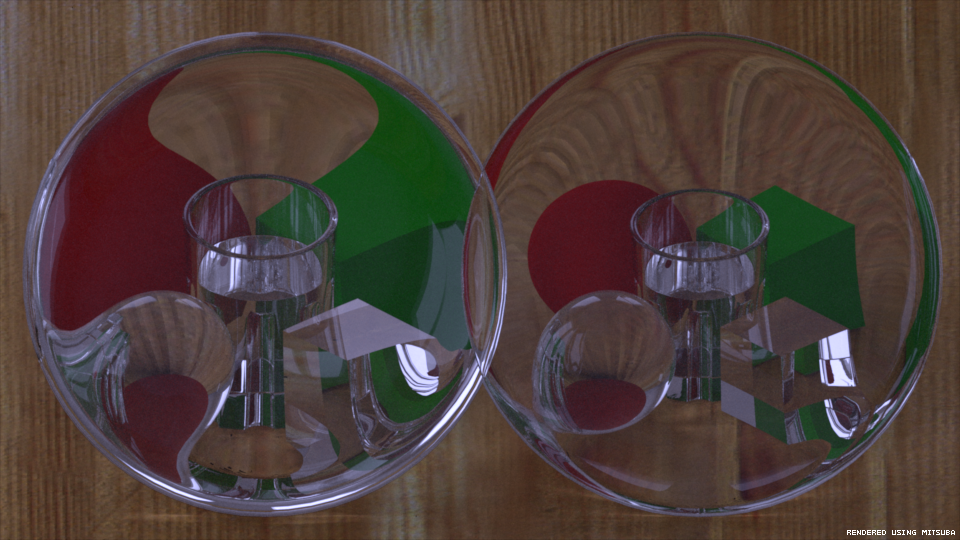}
	\caption{Lenses with different thickness depicting the effect such material and shape convey. Rendered with \gls{PSSMLT} and \emph{Independent sampler} with 1024 \gls{spp}.}
	\label{fig:lenses_differentroundness}
\end{figure}

Figure \ref{fig:lenses_differentroundness} depicts the effect of glass lenses with different thicknesses. The left one is a bit thicker than the right one, which results in a different magnification and focal point. Previously we have seen the effect in the shape and form of caustics. This time, a more direct approach was taken to show how lenses manipulate incoming light. 

The thicker lens to the left appears to magnify the objects in the back more than the right one. Considering the graphical interpretation of light passing through a lens in Figure \ref{fig:sphere_caustics}, this means that it depends on the shape, the thickness, and the curvature of the lens, how much the light will be refracted. This leads to the effect that images in thicker lenses appear bigger in size. It is recognizable that the objects placed along the borders of the lenses appear deformed. The thicker the lens, the more distorted objects become around the borders.

\subsection{Scene -- Double mirrors}\label{section:double_mirrors}
\begin{figure}[!htbp]
	\centering
	\includegraphics[width=\linewidth]{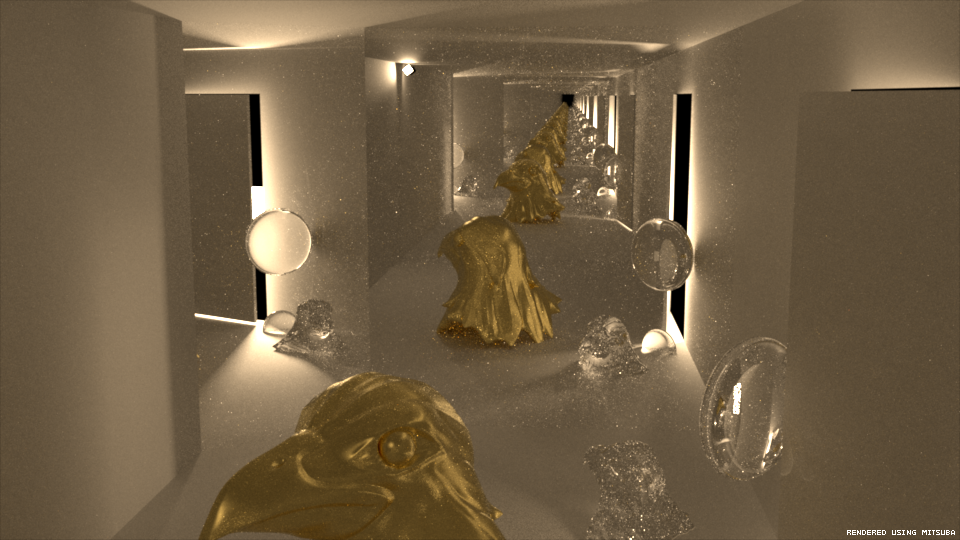}
	\caption{Multiple mirrors showing the impact of the bounces parameter for \emph{integrator}s. Rendered with \gls{PSSMLT} and \emph{Independent sampler} with 1024 \gls{spp}.}
	\label{fig:mirrormirror_onthewall} %
\end{figure}

The scene in Figure \ref{fig:mirrormirror_onthewall} was primarily designed to show the effect of the path depth parameter in the \emph{mitsuba} \emph{integrator}s. The path depth parameter is a general parameter that can be adjusted for any \emph{integrator}. The amount of the bounces control how many ray intersections happen per pixel (path depth). This can be observed in the rendering, the scene was rendered using a depth of 32 to show the effect of maximal path depth. This fact is noticeable when looking at the tiny black shaded square located in the mirror towards the top of the rendering. This indicates that the algorithm stopped computing and no further light is processed here. If the path depth is high enough or \emph{Russian Roulette} is employed, the unfilled area in the mirror would eventually get fully resolved. 

\emph{Russian Roulette} is a \emph{Monte Carlo} technique used for reducing variance originally introduced by Kirk et al. \cite{Kirk90}. Basically it stops tracing rays if there is less than a minimum of contribution to the pixel value. \cite{PharrEtAl2016}

Additionally, a high polygon object with a gold material preset has been placed to increase the complexity of the scene. Furthermore, two convex lenses are placed around the door to increase light complexity. The room itself consists of a door that is slightly opened. Outside the room, an area light is located to emit light towards the door crack as well as an area light inside the room which can be seen in the first mirrored instance. Usually, light coming from a tiny crack is difficult to catch for certain \emph{integrator}s because randomly shooting a ray towards the crack and light source is rare (\emph{Path tracer} has difficulties resolving such lighting situation). The rough diffuse material of the room increases the complexity of the scatterings as well.

\subsection{Scene -- Glass Prism}\label{section:glass_prism}
\begin{figure}[!htbp]
	\centering
	\includegraphics[width=\linewidth]{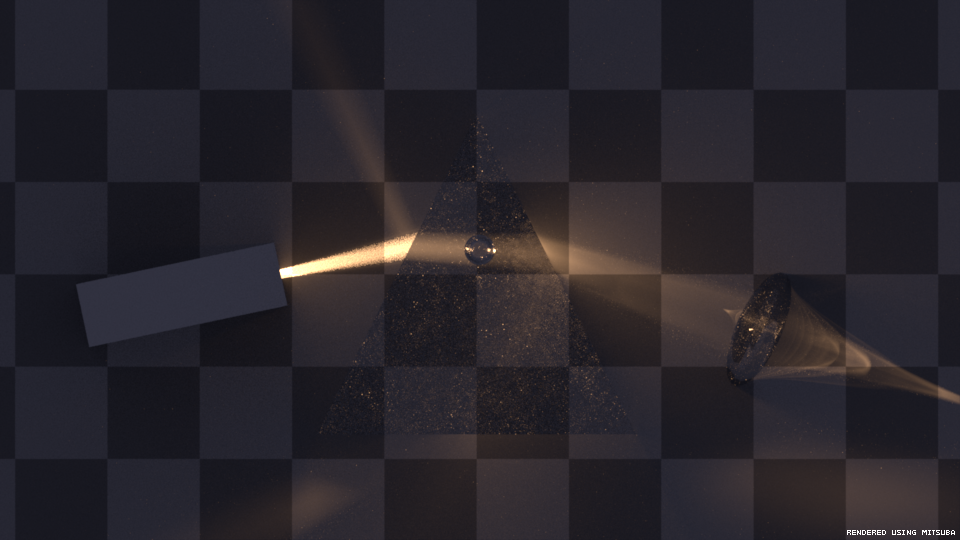}
	\caption{Glass prism which has interesting light reflections and transmissions. Rendered with \gls{PSSMLT} and \emph{Independent sampler} with 1024 \gls{spp}.}
	\label{fig:prisma} %
\end{figure}

Figure \ref{fig:prisma} shows a scene set up to describe the behavior of light falling onto a glass prism. The initial idea was to evaluate the correctness of light being refracted and reflected depending on the wavelengths of physical light. Although \emph{mitsuba} has an option to involve physical light with a discrete set of the wavelength spectrum, this paper and specifically this scene does not observe the behavior of different wavelengths, because the application would need some parameters changed and recompiled to enable this property. Furthermore, a vertical row of spheres is placed inside the prism to produce additional light reflection and refraction. It may appear as if the light rays pass through some \emph{participating medium}, but the illuminated areas on the ground simply come from the direct and indirect light. At the right hand side of the rendering there is a convex glass lens stuck in the ground, which collects the incoming light and throws a variety of caustics.

Regardless, interesting refractions and reflections of light can be observed in the scene. Like in the previous lens test scene, a light with high intensity was placed on the left side. It is also surrounded by geometry that helps the light shine in a narrower cone than usual. It is noticeable that one portion of the light gets reflected on the first intersection with the prism surface and the other portion gets transmitted through the glass material. Also mind the direction the whole bundle of light rays take when passing through the prism. The glass material has a typical index of refraction of 1.49, which is around an index of crown glass. The arrangement of the spheres in the middle of the prism have a focusing effect on the rays. The light bundle exiting the prism would normally have a lower intensity, since the spheres bundle the rays in the prism, the light appears brighter. Furthermore, it can be slightly seen that the glass spheres cast the typical sphere caustics. Considerably interesting are the caustics thrown from the lens on the right side.

\subsection{Scene -- Shadows of different light sources}\label{section:shadows_of_different_light_sources}
\begin{figure}[!htbp]
	\centering
	\includegraphics[width=\linewidth]{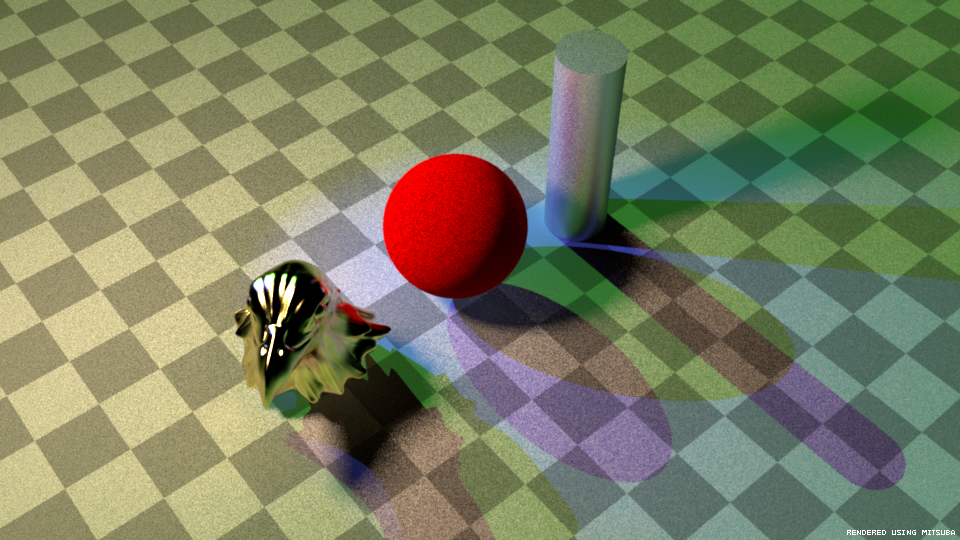}
	\caption{Light setup showing the different shadows cast from different light sources. Rendered with \gls{PSSMLT} and \emph{Independent sampler} with 1024 \gls{spp}.}
	\label{fig:softshadows} %
\end{figure}

The scene in Figure \ref{fig:softshadows} is designed to show the differences between the most common light sources in \emph{mitsuba}. The first one is a common area light in form of a sphere, the middle one is a spot light, and the last one is a directional light typically for the sun. 

Because of the properties of area lights defined in \emph{mitsuba}, this type of light source leads to objects casting soft shadows \cite{MitsubaDoc2014}. This is noticeable when observing the darkest portions of the shadows of the objects. The sharpness of the shadow boundary is dependent on the distance to the casting object. Close to the object the shadow boundary is sharp and further away it is more blurred. The same behavior can be seen for the other objects, whereas the area light shadow for the cylinder is barely visible because of the other lights overlapping it.

The spotlight is defined to have a certain linear falloff for the intensity \cite{MitsubaDoc2014}, however since the light emits from a single point in space, this type of light source does not cast soft shadows. This can be seen in the rendering when observing the blue light falling onto the objects. The shadows construct a perspectively distorted shape but the sharpness of the shadows remains constant.

The third light (green tint) is a direct light emitter, as defined in \emph{mitsuba} \cite{MitsubaDoc2014}. Typically, it simulates the sun like it exists in reality, which emits light rays in a specific direction. The common effect of a light source like the sun can be seen in the rendering by observing the shadows that the green light casts. Particularly, the shadows from the sphere and the cylinder appear to have a constant shape over distance. Soft shadows do not occur for this type of light source either.

Another interesting aspect is the difference of the materials of the objects. Whereas the complex object left has almost no grain, the red diffuse sphere and the rough metal cylinder exhibit a large amount of it as well as the shadows. This is because of the complexity of the surface material. The complex object has a smooth material, which makes it easier to render, whereas the red sphere and the cylinder have a diffuse material with moderate roughness that scatters the light to a greater extent.

\subsection{Scene -- Light from slightly opened door}\label{section:light_from_slightly_opened_door}
\begin{figure}[!htbp]
	\centering
	\includegraphics[width=\linewidth]{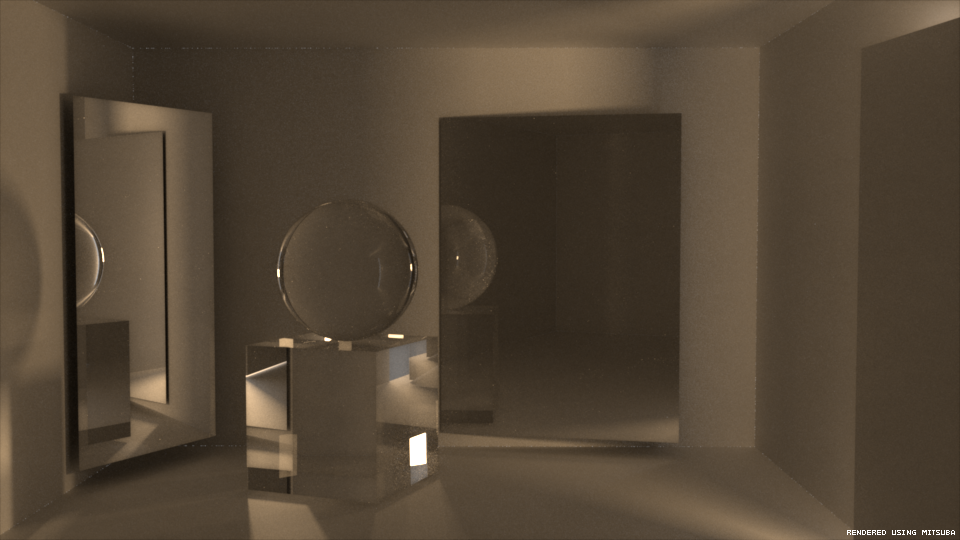}
	\caption{Light coming from a tiny crack of the door resulting in several light scattering events. Rendered with \gls{PSSMLT} and \emph{Independent sampler} with 1024 \gls{spp}.}
	\label{fig:softshadows_lightfromdoor} %
\end{figure}

The scene in Figure \ref{fig:softshadows_lightfromdoor} was specifically made to test the behavior of light coming from a door that is almost closed. The idea behind it was to observe a variety of soft shadows being cast all over the scene and to have a relatively simple setup of materials that simultaneously lead to a quick convergence and a visually pleasing result. The glass and mirror materials in the scene do not have any roughness, this is to decrease complexity overall and to put the focus on the light distribution through the door. The scene was rendered with the same amount of \gls{spp} (1024) as most of the other scenes, yet the result is already considerably good compared to other renderings. This is also because of the perfectly smooth glass and mirror surfaces. Several grainy portions can be observed on the right side of the cube where the light source from the door appears and on the back side of the sphere when looking at the right mirror. Interesting to observe are the cube shadows cast on the left wall on the left hand side of the rendering.

\subsection{Scene -- Light from outside}\label{section:light_from_outside}
\begin{figure}[!htbp]
	\centering
	\includegraphics[width=\linewidth]{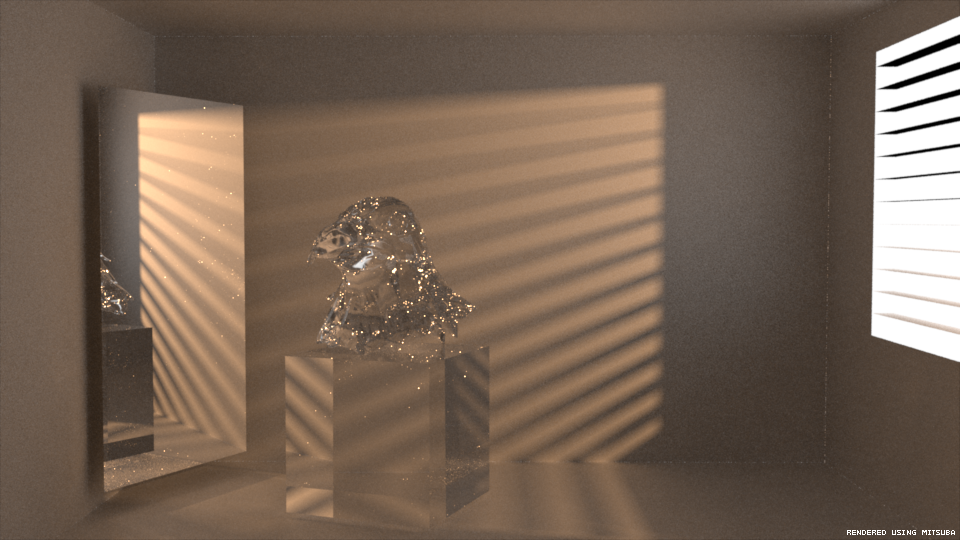}
	\caption{Light coming from outside and scattered through a simulated window shutter causing soft shadows. Rendered with \gls{PSSMLT} and \emph{Independent sampler} with 1024 \gls{spp}.}
	\label{fig:softshadows_lightfromoutside} %
\end{figure}

The scene in Figure \ref{fig:softshadows_lightfromoutside} is, similarly to previous scene, designed to show soft shadows and glass materials. This time the light comes from a window involving a shutter object that breaks and scatters the incoming rays. Additionally to an area light in form of a plane placed outside the window, the world environment is set to have a constant monochrome light. This can be seen when looking at the right-hand side around the window. The bright brownish light that appears on the back wall comes from the area light source. 

The effect of the decreased sharpness of soft shadows over distance can be distinctly seen when looking at the light projections at the back wall. Apart from the shadows getting more blurred, the further the light travels also the intensity of the reflected light gradually decreases. It is noteworthy that the light intensity in the mirror on the left side is higher than the intensity directly on the back wall. This happens because of the specularity of the room. The roughness is considerably high but is not independent from the point of view. The light reflecting from the diffuse back wall gets reflected by the mirror, which eventually reflects it back to the camera with a higher intensity. This is due to the angle towards the light source and therefore an angle that is closer around the perfect reflection angle of the wall.

A high polygon model is located on top of the cube. It is noteworthy that the graininess of the complex model differs considerably compared to the glass cube. There are more reflections outside and inside the high polygon object and therefore the scene is more difficult to render. Compared to the scene before (see Figure \ref{fig:softshadows_lightfromdoor}), this one has also been rendered with \gls{PSSMLT} with 1024 samples per pixel. Yet, the scene overall looks more grainy also because of the higher light intensity coming from outside.

\subsection{Scene -- \gls{SSS} and realistic light setup}\label{section:sss_and_realistic_light_setup}
\begin{figure}[!htbp]
	\centering
	\includegraphics[width=\linewidth]{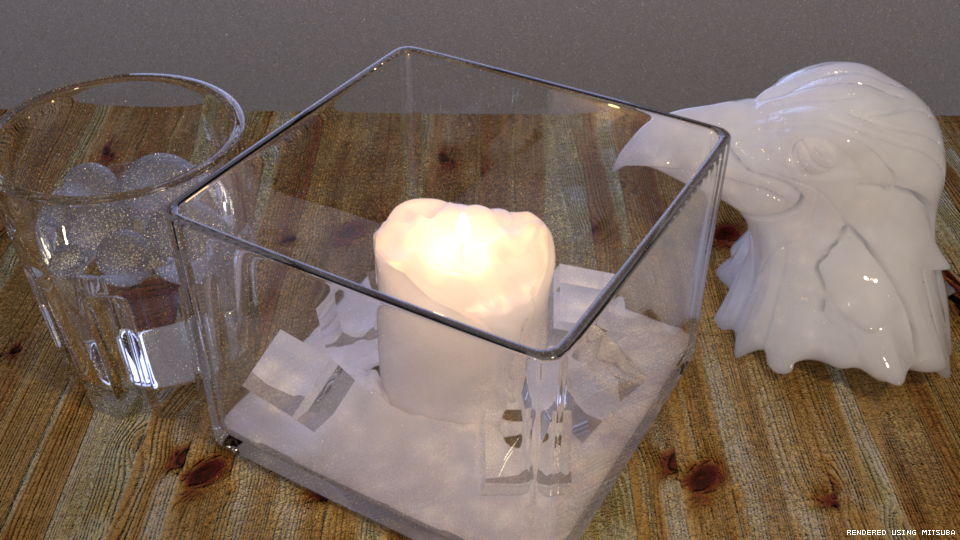}
	\caption{Scene with large \gls{SSS} portions and more realistic light setup. Rendered with \emph{Path tracer} and \emph{Independent sampler} with 1024 \gls{spp}.}
	\label{fig:sss_caustics} %
\end{figure}

The scene in Figure \ref{fig:sss_caustics} was specifically designed to show behavior of large portions of materials with \gls{SSS} in a more realistic light setting. Therefore, two high polygon objects are placed in the scene, one is a simulated wax candle and the other is the known high polygon object from previous scenes. The candle wick is lit and supports to show the impact of light on the subsurface material. Typically for candles, the light goes through the top of the lit surface and gradually decreases towards the bottom. 

The light source coming from outside is a constant environment tinted slightly blueish. This should simulate indirect environment lighting. Additionally to the light from outside, a lamp in the room was placed above the objects to give the scene more depth and several light scatterings.

Hard to render are the glass spheres located inside the water glass. They have a glass material with a moderate roughness, and therefore the graininess for this portion of the rendering is considerably high compared to the rest of the image. 

\subsection{Scene -- Complex \gls{SSS} and rough glass surfaces}\label{section:complex_sss_and_rough_glass_surfaces}
\begin{figure}[!htbp]
	\centering
	\includegraphics[width=\linewidth]{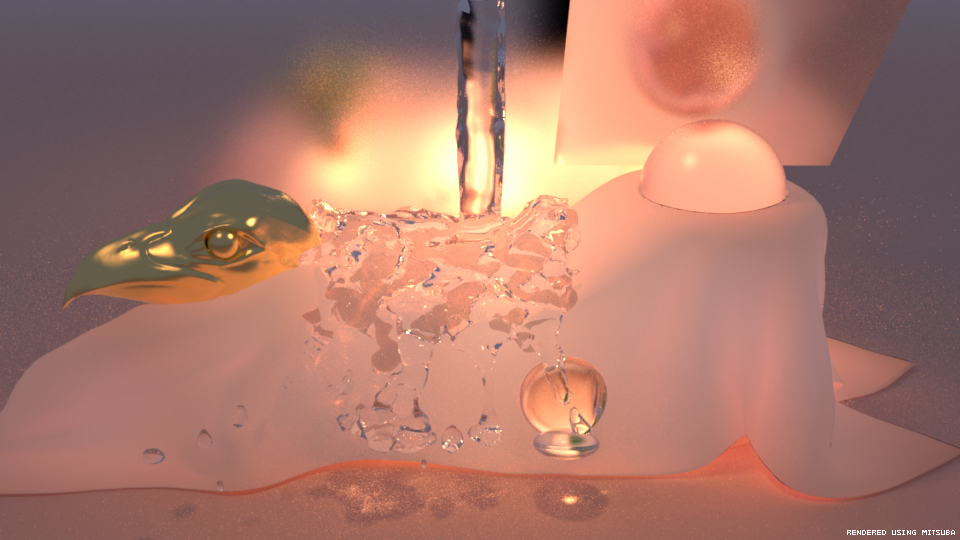}
	\caption{Complex portion of subsurface material in conjunction with rough glass surfaces. Rendered with \emph{Extended volumetric path tracer} and \emph{Independent sampler} with 1024 \gls{spp}.}
	\label{fig:sss_clothcover} %
\end{figure}

Figure \ref{fig:sss_clothcover} shows a scene with several different materials put together. Generally, this scene was designed with a more artistic and more complex approach in mind. The cloth blanket that covers objects underneath has an \gls{SSS} material applied to simulate more realistic clothing. The high polygon object to the left has a gold material applied. Underneath the blanket, in the middle of the scene, there is a glass box with a red diffuse sphere inside and the object to the right is a water glass. Behind in the back of the scene, a mirror with moderate roughness distributes the light that shines onto the blanket from behind. Another light is placed above to allow the water and lenses to throw visible caustics. Additionally to these objects a water outflow was put above the scene mainly to let the water throw caustics on the ground. Another rough glass sphere is placed on top of the glass to the right to emphasize light scattering inside the blanket. In that respect, it is also noticeable that the bottom of the blanket is highlighted from the scattered light inside.

The scene was rendered with the \emph{Extended volumetric path tracer} (volpath) \cite{MitsubaDoc2014} \emph{integrator}. Also mind the relatively sharp and smooth portions of the rendering that do not belong to glass caustics. This is a well known problem for volpath. Yet, for 1024 samples per pixel, the caustics on the bottom of the image coming from the water drops and the lens are considerably good but still grainy in comparison when rendered with an \emph{integrator} like \gls{PSSMLT}. It is noticeable for the blanket that there are several cloth folds that generally receive less light from the internal scattering events and therefore convey a better perception of depth. Also interesting is to show \gls{SSS} with parameters that are typical for cloth and how little is visible through the highly scattering material.

\subsection{Scene -- Water caustics and broad light distribution}\label{section:water_caustics_and_broad_light_distribution}
\begin{figure}[!htbp]
	\centering
	\includegraphics[width=\linewidth]{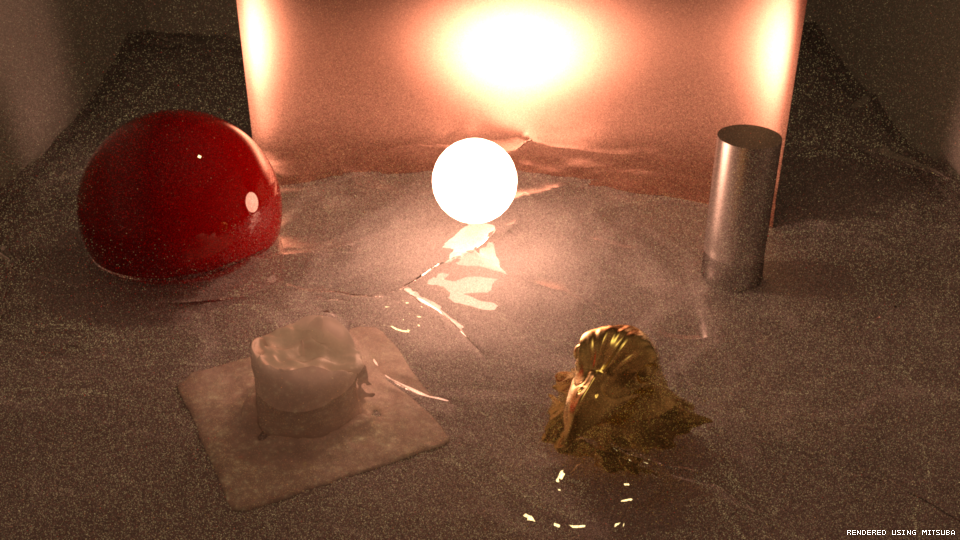}
	\caption{Room filled with water and objects with a sphere light in the middle. Rendered with \emph{Extended volumetric path tracer} and \emph{Independent sampler} with 1024 \gls{spp}.}
	\label{fig:underwater_light} %
\end{figure}

The scene in Figure \ref{fig:underwater_light} is a constellation of objects with different materials arranged around an area light in form of a sphere. Starting with the first object, a bent mirror with a slightly rough copper material is placed behind the sphere light to enable diffuse light scatterings. On the left side, a red diffuse sphere within a larger glass sphere is placed. To give the scene a more complex material setup, a wax candle with an \gls{SSS} material was added. Besides the high polygon wax candle, the raven head was added as well and has been applied with a gold material. The last object is a simple cylinder with a metal texture with moderate roughness.

Additionally, a water mesh fills the room so about half of the height of the objects are located inside. It is noticeable that for the red sphere, the reflection of the light source is not as intense. This comes from a light portion getting reflected by the outer glass sphere and the other portion hitting the red diffuse material. The roughness of the bent mirror ensures that there are many light scattering events that take place over the scene. This increases complexity in terms of rendering time and visual quality. The volpath \emph{integrator} has been used to render the image. The issue is that the scene has an \gls{SSS} material and simultaneously there are many caustics that need to be rendered. This leads to the problem that \emph{integrator}s like \gls{PSSMLT} would not work because of the implementation in \emph{mitsuba}. The deficit of volpath not being able to handle caustics well, is noticeable when looking at the darker portions of the rendering. The perceivable variance is very high and the image as a whole looks very grainy.

\section{Evaluation}

This chapter deals with the evaluation of the scenes that are described in Chapter 2. Depending on the scene, the \emph{integrator}s achieve considerably different results within the same render time. Each scene evaluation will contain five (four in special cases) render results from a selection of the most relevant \emph{integrator}s. The algorithms partially sample light very differently and comparing on the basis of equal \gls{spp} would hide the sampling overhead of the different algorithms. This is the reason why the results from the renderings are compared with render time. Perfectly accurate consensus in render time is not possible, but the differences in time are around plus/minus ten seconds.

\subsection{Integrators} 

The following will give brief descriptions about the \emph{integrator}s as well as the abbreviations that are used in the evaluation.
 
\paragraph{\gls{PT}}

The \emph{Path tracer} is a basic path tracer implementation that shoots random rays from the camera into the scene. It is a good choice when the scene contains only simple lighting with no obstacles along the light paths.  \cite{MitsubaDoc2014}

It uses \emph{Multiple Importance Sampling} to reduce variance in the image. \emph{Multiple Importance Sampling} is a technique in \emph{Monte Carlo}-sampling that uses the idea of taking samples from multiple different probability density functions $p(x)$ (like described in Section \ref{subsubsection:importance_sampling}), in the hope that one of the distributions resembles the shape of the integrand. \cite{PharrEtAl2016}

\paragraph{\gls{VOLPATH_SIMPLE}} 

The \emph{Simple volumetric path tracer} implements a basic volumetric path tracer that is used to render volumes in scenes (\emph{participating media}, \gls{SSS} materials). It does not make use of \emph{Multiple Importance Sampling}. \cite{MitsubaDoc2014}

\paragraph{\gls{VOLPATH}}

This \emph{integrator} implements a volumetric path tracer and it also uses \emph{Multiple Importance Sampling}. In respect of surfaces, it behaves like the standard \emph{Path tracer} \cite{MitsubaDoc2014}

\paragraph{\gls{BDPT}}

The \gls{BDPT} is implemented as proposed by Veach and Guibas \cite{Veach95}, in conjunction with \emph{Multiple Importance Sampling}. The basic idea is to start two random ray paths in one step. One ray starts from the camera and one from an emitter as described in Section \ref{subsubsection:bidirectional_sampling}. \gls{BDPT} usually is considerably slower than \emph{Path tracing} for the same amount of samples but generally results in images with much less perceivable variance. \cite{MitsubaDoc2014}

\paragraph{\gls{PM}}

The \emph{Photon mapper} implementation is based on the idea proposed by Jensen \cite{Jensen96}. The calculation of light is separated into three classes. Diffuse, caustic, and volumetric maps are built separately. A ray tracing pass follows that estimates the radiance using the photon maps.

\paragraph{\gls{PPM}}

This \emph{integrator} implements \emph{Progressive photon mapping} by Hachisuka et al. \cite{HachisukaPPM}. It is based on \emph{Photon mapping} but it progressively calculates and shows the results as the passes continue indefinitely until eventually all variance vanishes.

\paragraph{\gls{SPPM}}

This \emph{integrator} implements \gls{SPPM} as proposed by Hachisuka et al. \cite{HachisukaSPPM}. It is an extension of \emph{Progressive photon mapping} that has better performance when it comes to motion blur, depth of field or glossy reflections.

\glslocalreset{PSSMLT}
\paragraph{\gls{PSSMLT}}
This rendering technique was developed by Kelemen et al. \cite{Kelemen-2001-METR}, based on \emph{Markov Chain Monte Carlo Integration} \cite{Hastings_MCMC70}. Usually, algorithms like \emph{path tracing} compute images by randomly shooting light paths into the scene. \gls{PSSMLT} benefits from the fact that it tries to find most relevant light paths. If a relevant path is discovered also neighbored paths are evaluated and involved in the calculation. Generally, this improves the render time considerably and complex light situations give a better result. \cite{MitsubaDoc2014}

\paragraph{\gls{MLT}}

\gls{MLT} is similar to \gls{PSSMLT}. Both \emph{integrator}s search for light paths that carry a high amount of light energy. \gls{PSSMLT} does this by using another rendering technique. \gls{MLT} instead works directly on the light carrying paths and therefore has more information available. This allows directed search for certain classes of light paths. \cite{MitsubaDoc2014}

\paragraph{\gls{ERPT}}

This \emph{integrator} by Cline et al. \cite{ClineERPT} uses \emph{Path tracing} in conjunction with techniques of \emph{Metropolis Light Transport}. With the use of a standard bidirectional path tracer, a set of seed paths is generated. An \gls{MLT} Markov Chain is then started for each path that redistributes the energy of the samples over a larger area.

\paragraph{\gls{IRRCACHE}}

\emph{Irradiance caching} is a \emph{meta-integrator} that works on top of other \emph{integrator}s. It was developed by Ward and Heckbert \cite{WardIrrcache}. It calculates and caches irradiance information at several scene locations and fills the other locations through interpolation. This can be helpful if results would otherwise look blotchy (often used with photon mapping).

\subsection{Water glass}

\begin{figure*}[!htbp]
	\centering
	\includegraphics[width=\linewidth]{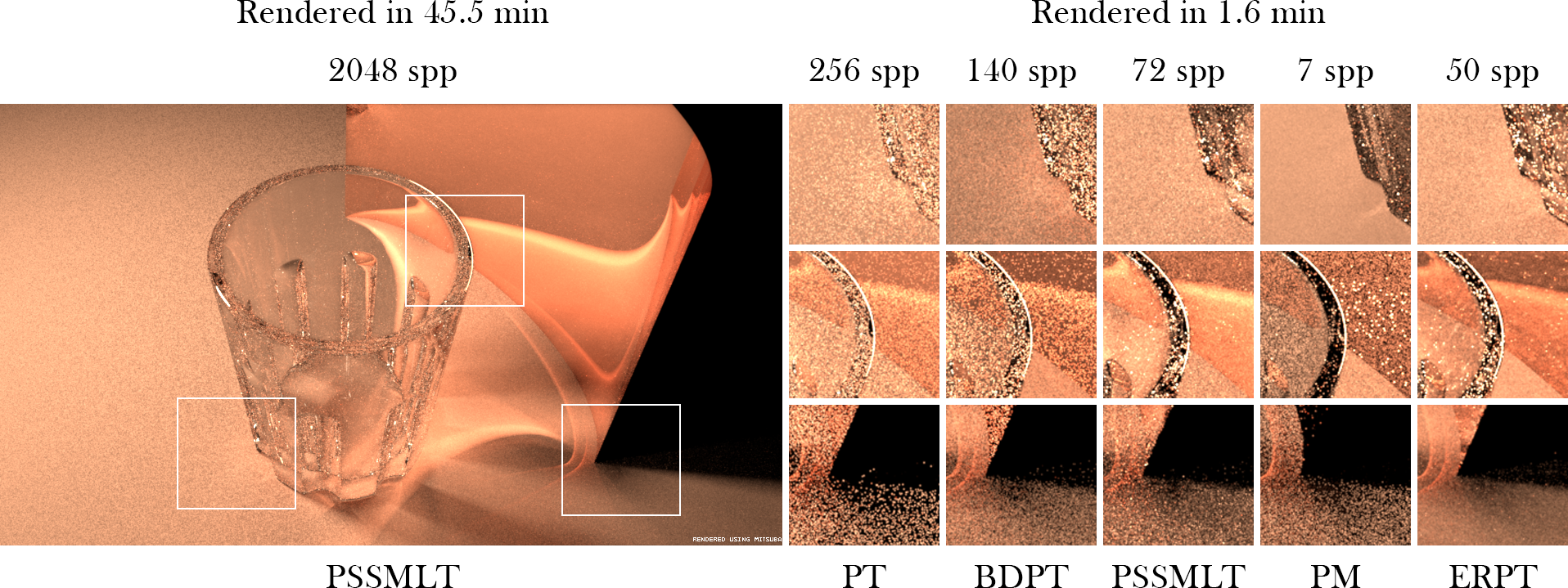}
	\caption{Reference rendered with \gls{PSSMLT} (Independent Sampler) at 2048 \gls{spp}. Comparison images (Independent Sampler) rendered in 1.6 minutes.}
	\label{fig:caustic_direct_light_evaluation} %
\end{figure*}

The reference in Figure \ref{fig:caustic_direct_light_evaluation} was rendered with \gls{PSSMLT} with 2048 samples per pixel (\gls{spp}). The result of the \emph{path tracer} (\gls{PT}) \emph{integrator} was used as a benchmark in terms of render time. The images where rendered in about 1.6 minutes. Concerning soft shadows (bottom row), \gls{PT} had difficulties delivering good results. \gls{BDPT} resulted in considerably lower perceived variance. \gls{PSSMLT} and \gls{PM} also had difficulties rendering soft shadows and resulted in a lower visual quality than \gls{BDPT}. The best results were achieved by \gls{ERPT}.

The glass portions (Figure \ref{fig:caustic_direct_light_evaluation} middle and top row) of the scene are hard to render. Both \gls{PT} and \gls{BDPT} had troubles resolving these parts. \gls{PSSMLT} and \gls{ERPT} achieved the best results while \gls{PM} delivered a result that is too dark overall. However, \gls{PM} resolved the caustic on left side of the glass (on the ground) better than the other \emph{integrator}s. This can be seen in Figure \ref{fig:caustic_direct_light_evaluation} \gls{PM} top row, it shows considerably less perceivable variance than the other results.

As a conclusion, it can be stated that there is no \emph{integrator} that performed well in all aspects. While \gls{PM} had better results on portions of the ground, the glass was considerably better for \gls{PSSMLT} and \gls{ERPT}. Both \gls{PT} and \gls{BDPT} had difficulties in most cases.

\subsection{Glass pendulum}
\begin{figure*}[!htbp]
	\centering
	\includegraphics[width=\linewidth]{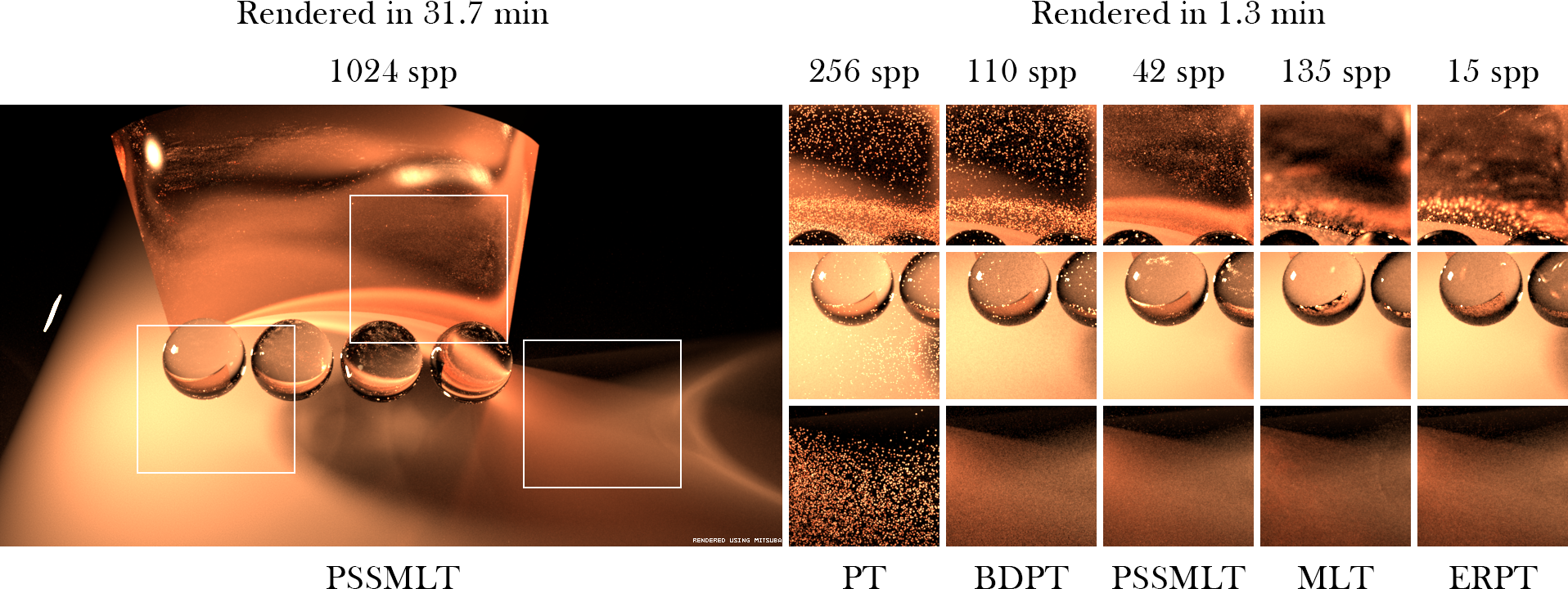}
	\caption{Reference rendered with \gls{PSSMLT} (Independent Sampler) at 1024 \gls{spp}. Comparison images (Independent Sampler) rendered in 1.3 minutes.}
	\label{fig:caustic_direct_light_pendulum_evaluation}
\end{figure*}

The reference in Figure \ref{fig:caustic_direct_light_pendulum_evaluation} was rendered with \gls{PSSMLT} with 1024 samples per pixel. All comparison images were rendered in 1.3 minutes. The \gls{PT} has problems resolving the caustics that are thrown by the spheres. Also the bent mirror in the background (which has little roughness) has a considerably high perceivable variance. While the \gls{BDPT} achieves good results on the portions of the ground (shadows and caustics) the bent mirror also has a high perceivable variance and is not visually better than the result from \gls{PT}.

\gls{PSSMLT} resolves all aspects better than the other \emph{integrator}s. \gls{MLT} and \gls{ERPT} resolve the ground similarly like \gls{PSSMLT} but have problems rendering the bottom part of the bent mirror (Figure \ref{fig:caustic_direct_light_pendulum_evaluation} top row). This can be seen as little light peaks all over the bottom of the mirror.

\subsection{Glass pendulum different shapes}
\begin{figure*}[!htbp]
	\centering
	\includegraphics[width=\linewidth]{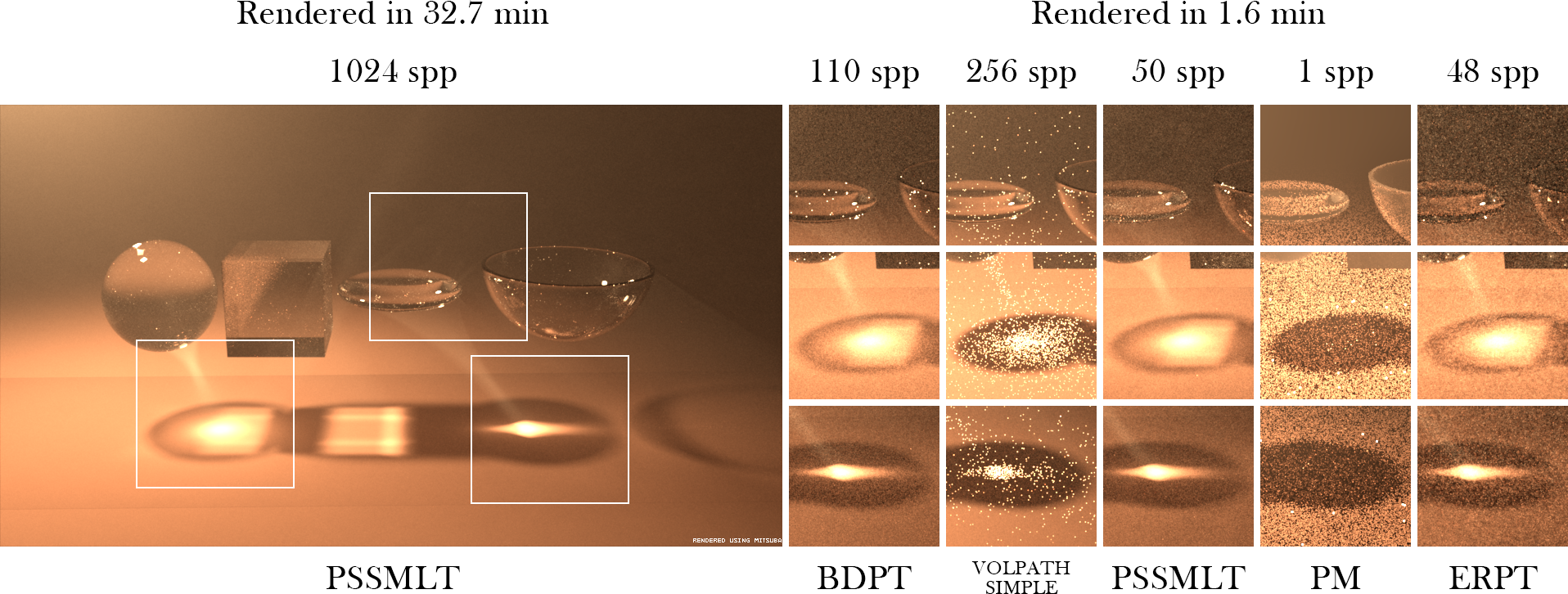}
	\caption{Reference rendered with \gls{PSSMLT} (Independent Sampler) at 1024 \gls{spp}. Comparison images (Independent Sampler) rendered in 1.6 minutes.}
	\label{fig:caustic_direct_light_pendulum_pm_evaluation}
\end{figure*}

The reference in Figure \ref{fig:caustic_direct_light_pendulum_pm_evaluation} was rendered with \gls{PSSMLT} with 1024 samples per pixel. The comparison images were rendered in 1.6 minutes. The \gls{BDPT} rendered the scene considerably well. If compared to the result from \gls{PSSMLT} (which achieved the visually best result), there is not much difference. Little drawbacks can be seen in terms of higher perceivable variance for the \gls{BDPT}.

\gls{VOLPATH_SIMPLE} did not achieve good results, the \emph{integrator} had issues finding the light carrying paths as well as light scattered inside the volume of the \emph{participating medium}. The \gls{PM} did not work well, it did not catch the caustics on the ground and it had problems resolving scattered light inside the volume. The results from \gls{ERPT} are behind \gls{PSSMLT} and \gls{BDPT} because the perceivable variance is even higher but it resolves the caustics and volume visually better than \gls{PM} and \gls{VOLPATH_SIMPLE}.

\subsection{Sphere and lenses}
\begin{figure*}[!htbp]
	\centering
	\includegraphics[width=\linewidth]{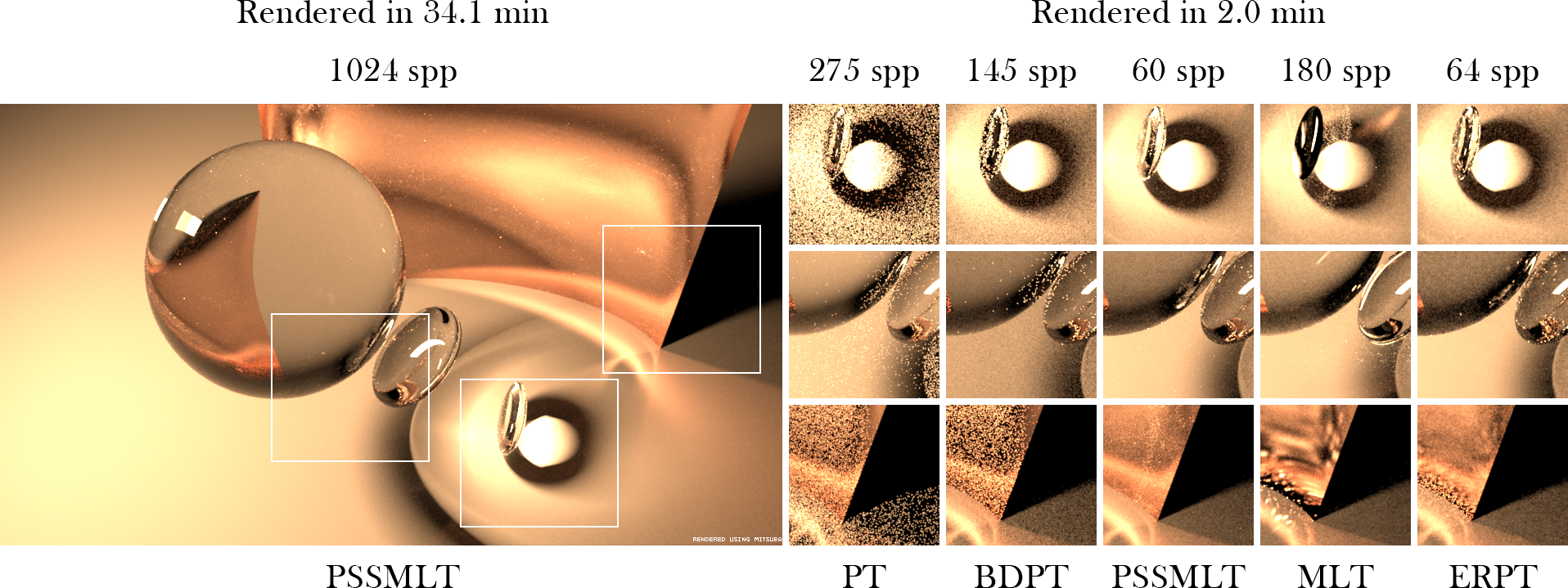}
	\caption{Reference rendered with \gls{PSSMLT} (Independent Sampler) at 1024 \gls{spp}. Comparison images (Independent Sampler) rendered in 2.0 minutes.}
	\label{fig:caustic_direct_light_sphere_evaluation} %
\end{figure*}

The reference in Figure \ref{fig:caustic_direct_light_sphere_evaluation} was rendered with \gls{PSSMLT} with 1024 samples per pixel. The comparison images were rendered in 2.0 minutes. Like in the previous scenes, the \gls{PT} has issues resolving caustics and in general light paths that are considerably complex. This results again in high perceivable variance all over the scene. The \gls{BDPT} achieves relatively smooth results on the ground portions but not for the bent mirror in the background (more grainy). 

\gls{PSSMLT} again achieves the visually best result among all \emph{integrator}s although \gls{MLT} resolves the lenses a bit better (Figure \ref{fig:caustic_direct_light_sphere_evaluation} \gls{MLT} middle row). \gls{MLT} has several light phenomena/caustics that are not present in the reference image. While \gls{ERPT} does a good job for the lenses and ground portions, it does not resolve the bottom of the mirror, neither does \gls{MLT}.

\subsection{Water caustics 1}
\begin{figure*}[!htbp]
	\centering
	\includegraphics[width=\linewidth]{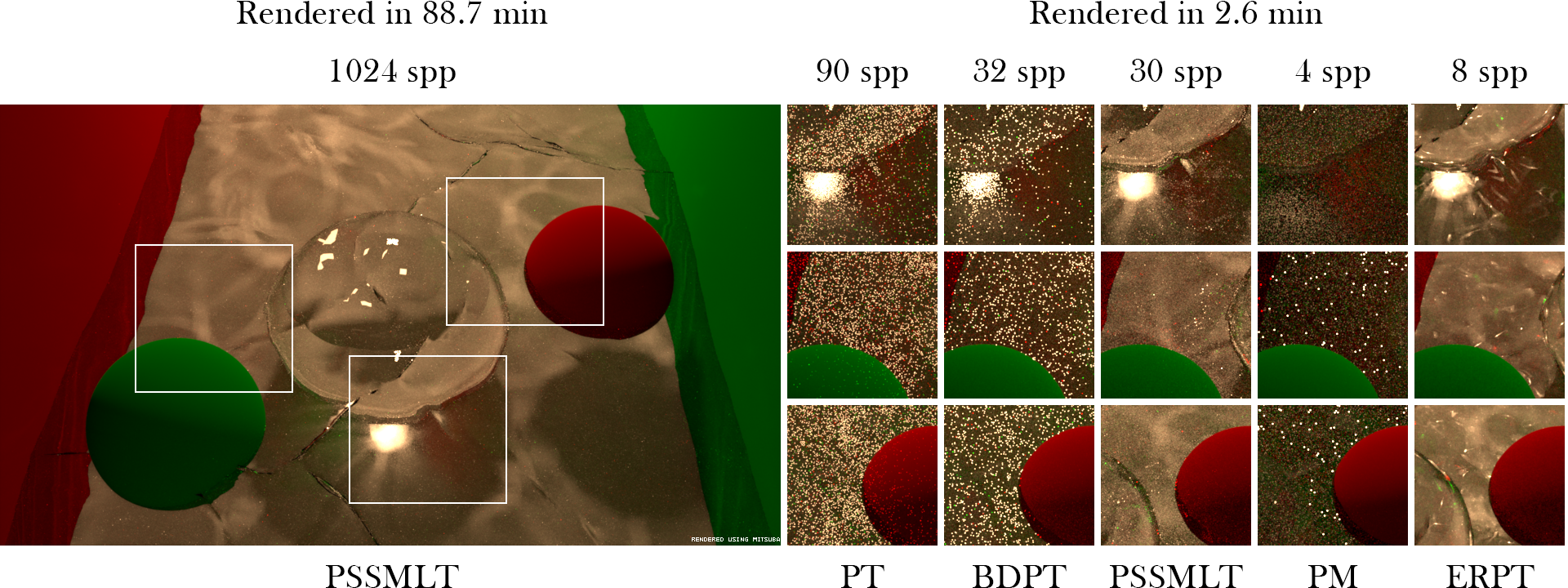}
	\caption{Reference rendered with \gls{PSSMLT} (Independent Sampler) at 1024 \gls{spp}. Comparison images (Independent Sampler) rendered in 2.6 minutes.}
	\label{fig:caustic_direct_light_waterwaves_evaluation} %
\end{figure*}

The reference in Figure \ref{fig:caustic_direct_light_waterwaves_evaluation} was rendered with \gls{PSSMLT} with 1024 samples per pixel. The comparison images were rendered in 2.6 minutes. The scene lighting in conjunction with the waves on the ground makes this scene difficult to render. The \gls{PT} achieves results with high perceivable variance. However, in comparison to the \gls{BDPT} it occurs to resolve light paths better. 

\gls{PSSMLT} has little issues and resolves the caustics relatively well. \gls{PM} results in a considerably darker scene when trying to catch the light and caustics from the water, all other parts (the diffuse spheres) are similar to the other results. \gls{ERPT} achieves the best result when it comes to caustics even though there are light artifacts all over the water, however, the caustics generally exhibit less perceivable variance than for \gls{PSSMLT}.

\subsection{Water caustics 2}
\begin{figure*}[!htbp]
	\centering
	\includegraphics[width=\linewidth]{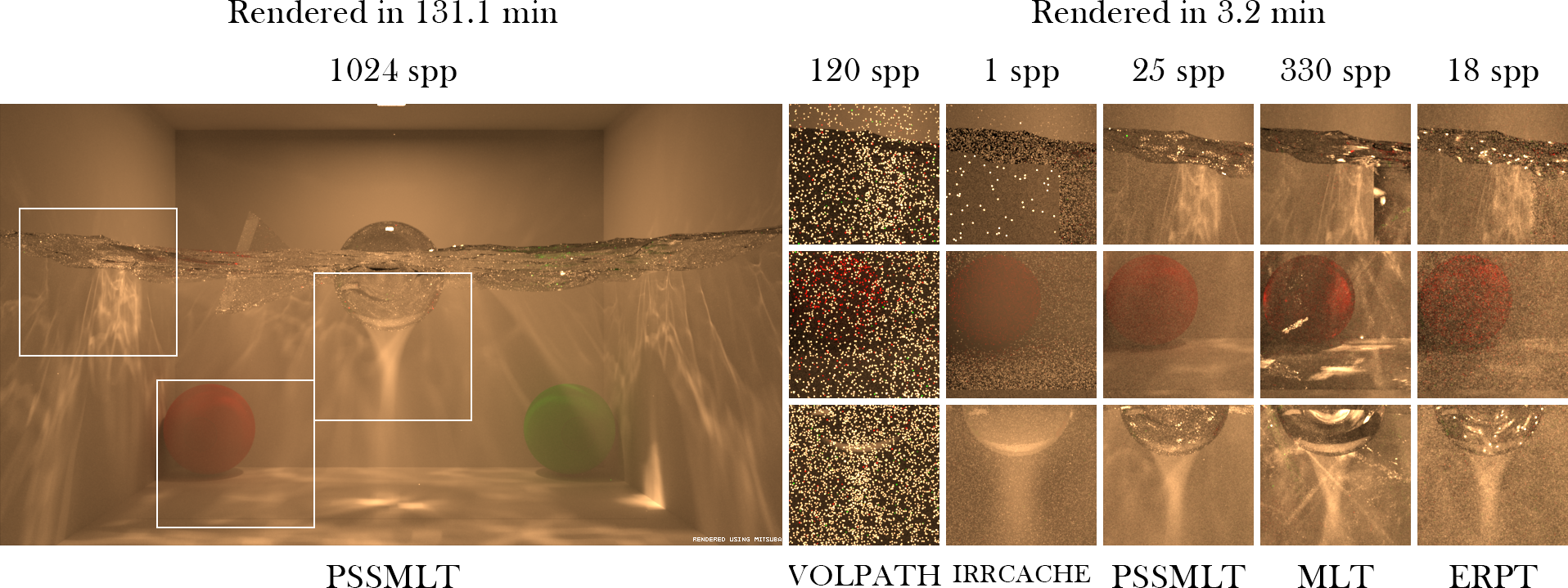}
	\caption{Reference rendered with \gls{PSSMLT} (Independent Sampler) at 1024 \gls{spp}. Comparison images (Independent Sampler) rendered in 3.2 minutes.}
	\label{fig:caustic_direct_light_waterwaves_pm_evaluation} %
\end{figure*}

The reference in Figure \ref{fig:caustic_direct_light_waterwaves_pm_evaluation} was rendered with \gls{PSSMLT} with 1024 samples per pixel. The comparison images were rendered in 3.2 minutes. The \gls{VOLPATH} \emph{integrator} did not achieve good results, the noise is very high. The \gls{PM} was used with \gls{IRRCACHE} but had problems resolving certain parts as well, especially for caustics.

\gls{PSSMLT} had no problems rendering the scene. The perceivable variance seems to be the lowest among all \emph{integrator}s. \gls{MLT} results in appropriate caustics as well on the one hand, on the other hand, it shows many individual light features all over the scene. Also the scene overall looks a bit darker than for the other \emph{integrator}s. \gls{ERPT} did not create any artifacts but the perceivable variance seems to be still higher than for \gls{PSSMLT} and \gls{MLT}.

\subsection{Color bleeding}
\begin{figure*}[!htbp]
	\centering
	\includegraphics[width=\linewidth]{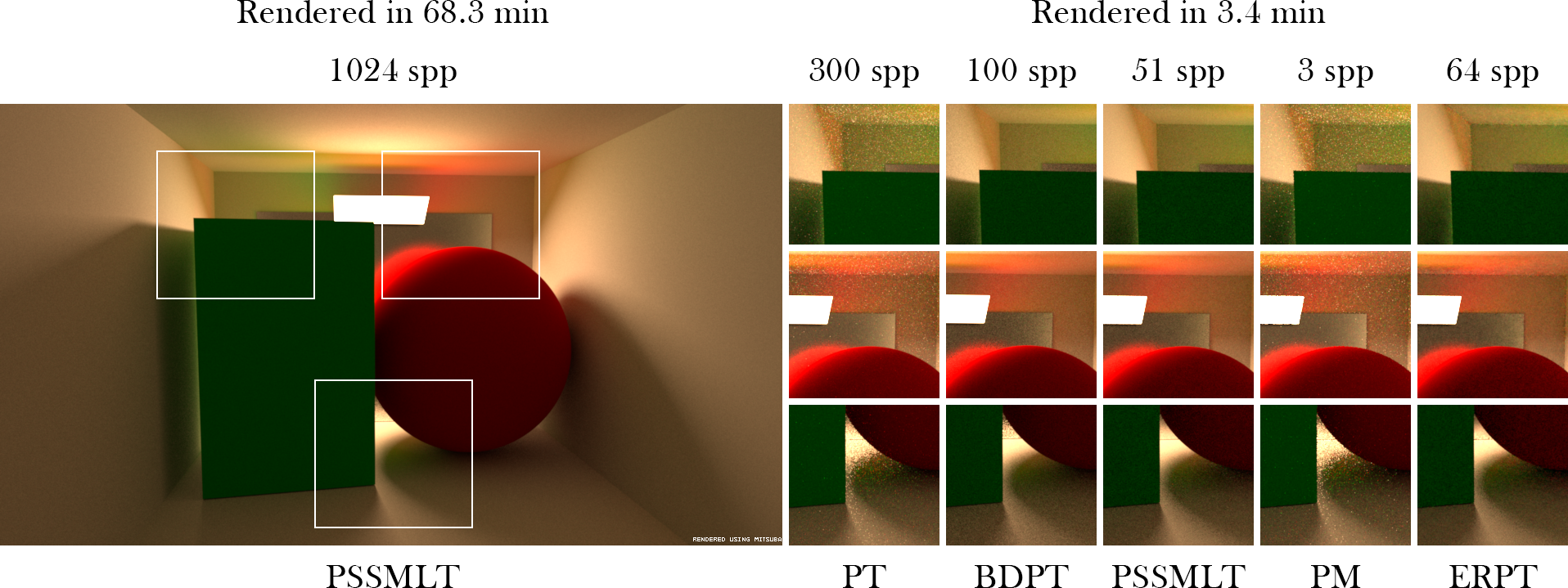}
	\caption{Reference rendered with \gls{PSSMLT} (Independent Sampler) at 1024 \gls{spp}. Comparison images (Independent Sampler) rendered in 3.4 minutes.}
	\label{fig:colorbleeding_evaluation} %
\end{figure*}

The reference in Figure \ref{fig:colorbleeding_evaluation} was rendered with \gls{PSSMLT} with 1024 samples per pixel. The  comparison images were rendered in 3.4 minutes. The \gls{PT} shows high perceivable variance in the resulting image. \gls{BDPT} as well as \gls{PSSMLT} delivered very similar results. \gls{BDPT} only had more issues resolving the mirror in the background that shows slightly higher perceivable variance than for \gls{PSSMLT} (Figure \ref{fig:colorbleeding_evaluation} mirror in bottom and mid row).

The \gls{PM} overall resulted in an image with high perceivable variance comparable to the result from the \gls{PT}. The \gls{ERPT} achieved good results as well and is similar to the results from \gls{BDPT} and \gls{PSSMLT}, however, the scene overall looks slightly blurry. To conclude, the \gls{BDPT} is probably preferable here because it has slight benefits in terms of memory usage compared to \gls{PSSMLT}.

\subsection{Smooth and rough glass}
\begin{figure*}[!htbp]
	\centering
	\includegraphics[width=\linewidth]{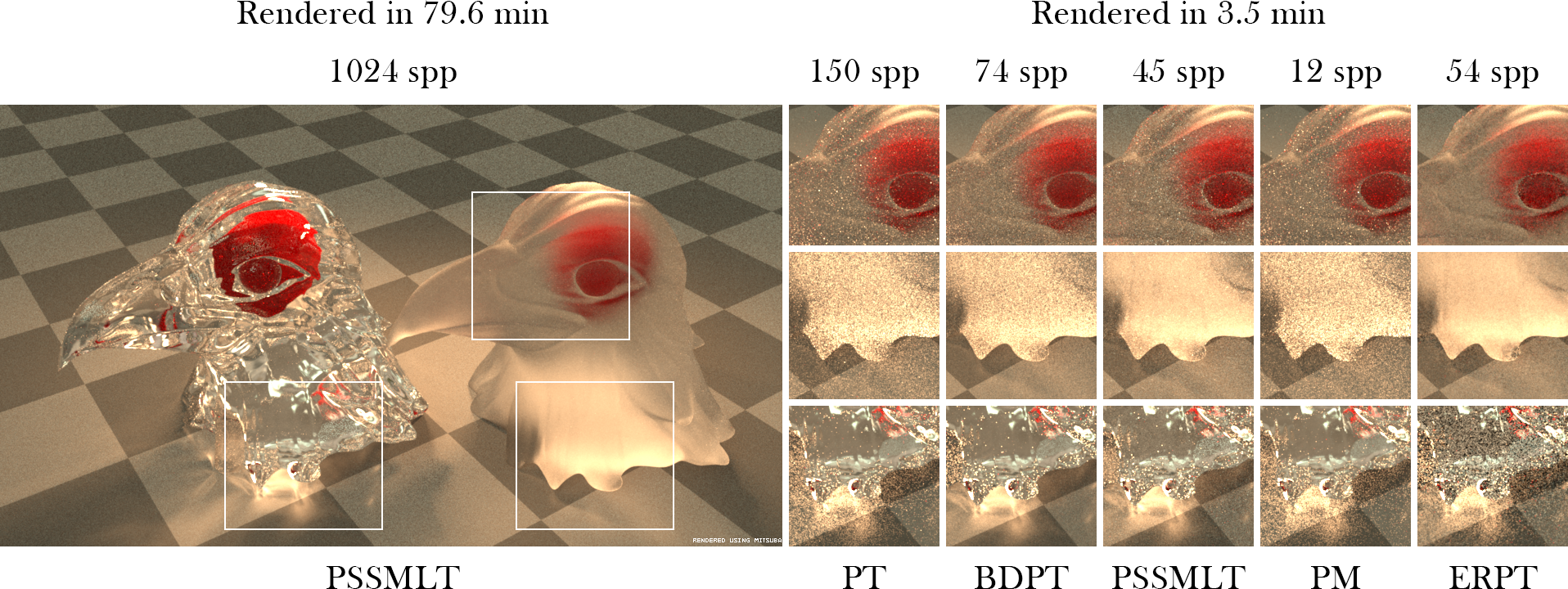}
	\caption{Reference rendered with \gls{PSSMLT} (Independent Sampler) at 1024 \gls{spp}. Comparison images (Independent Sampler) rendered in 3.5 minutes.}
	\label{fig:complex_glass_evaluation} %
\end{figure*}

The reference in Figure \ref{fig:complex_glass_evaluation} was rendered with \gls{PSSMLT} with 1024 samples per pixel. The comparison images were rendered in 3.5 minutes. The \gls{PT} had troubles rendering the caustics from the smooth glass object on the left side as well as from the rough glass object on the right side. In both cases, the perceivable variance is higher compared to other \emph{integrator}s. \gls{BDPT} did a considerably better job at rendering the caustics. However, it also has the same issues when it comes to rendering the interior lighting of the glass material and exhibits high perceivable variance, especially for the rough glass object.

\gls{PSSMLT} achieved good results overall. It does well on caustics and the rough glass parts. Yet, the \gls{BDPT} resolves the caustics a bit better (Figure \ref{fig:complex_glass_evaluation} bottom row). The \gls{PSSMLT} has a lower perceivable variance for the bottom part of the rough glass (Figure \ref{fig:complex_glass_evaluation} middle row) but the \gls{BDPT} has a lower perceivable variance for the top part of the rough glass (Figure \ref{fig:complex_glass_evaluation} top row). 

The \gls{PM} exhibits similar perceivable variance like the \gls{PT} and is not suitable for this scene. The rendering from the \gls{ERPT} has the lowest perceivable variance for the rough glass object but exhibits artifacts for the smooth glass object on the left side. It seems that \gls{PSSMLT} achieved the best result overall.

\subsection{\gls{SSS} complex}
\begin{figure*}[!htbp]
	\centering
	\includegraphics[width=\linewidth]{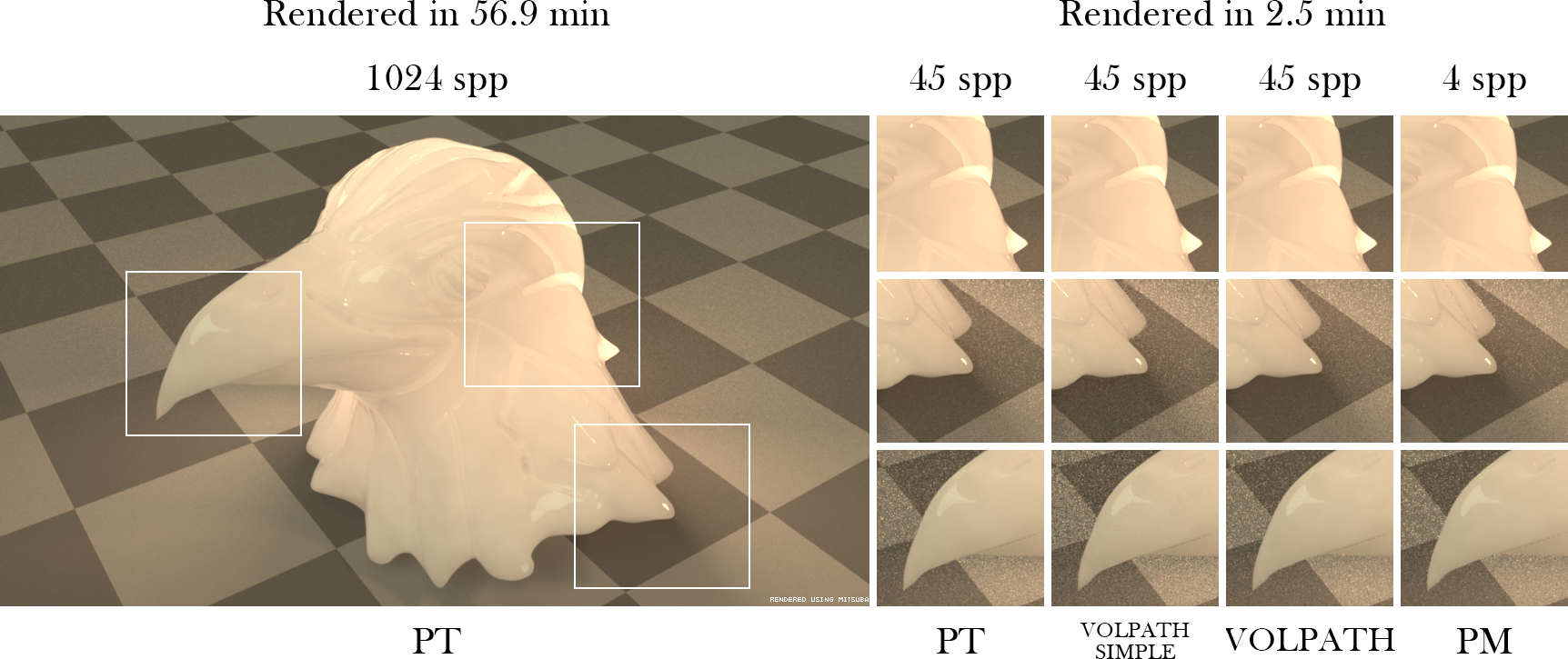}
	\caption{Reference rendered with \gls{PT} (Independent Sampler) at 1024 \gls{spp}. Comparison images (Independent Sampler) rendered in 2.5 minutes.}
	\label{fig:complex_sss_evaluation}
\end{figure*}

The reference in Figure \ref{fig:complex_glass_evaluation} was rendered with the \gls{PT} with 1024 samples per pixel. The comparison images were rendered in 2.5 minutes. Only \gls{PT}, \gls{VOLPATH}, \gls{VOLPATH_SIMPLE} and \gls{PM} are able to capture \gls{SSS} materials. The results from the different \emph{integrator}s are almost equal. \gls{PT}, \gls{VOLPATH_SIMPLE} and \gls{VOLPATH} achieve the same result in terms of visual quality and memory usage. \gls{PM} results in a very similar image but it uses slightly more memory for the calculations.

\subsection{\gls{SSS} complex and water}
\begin{figure*}[!htbp]
	\centering
	\includegraphics[width=\linewidth]{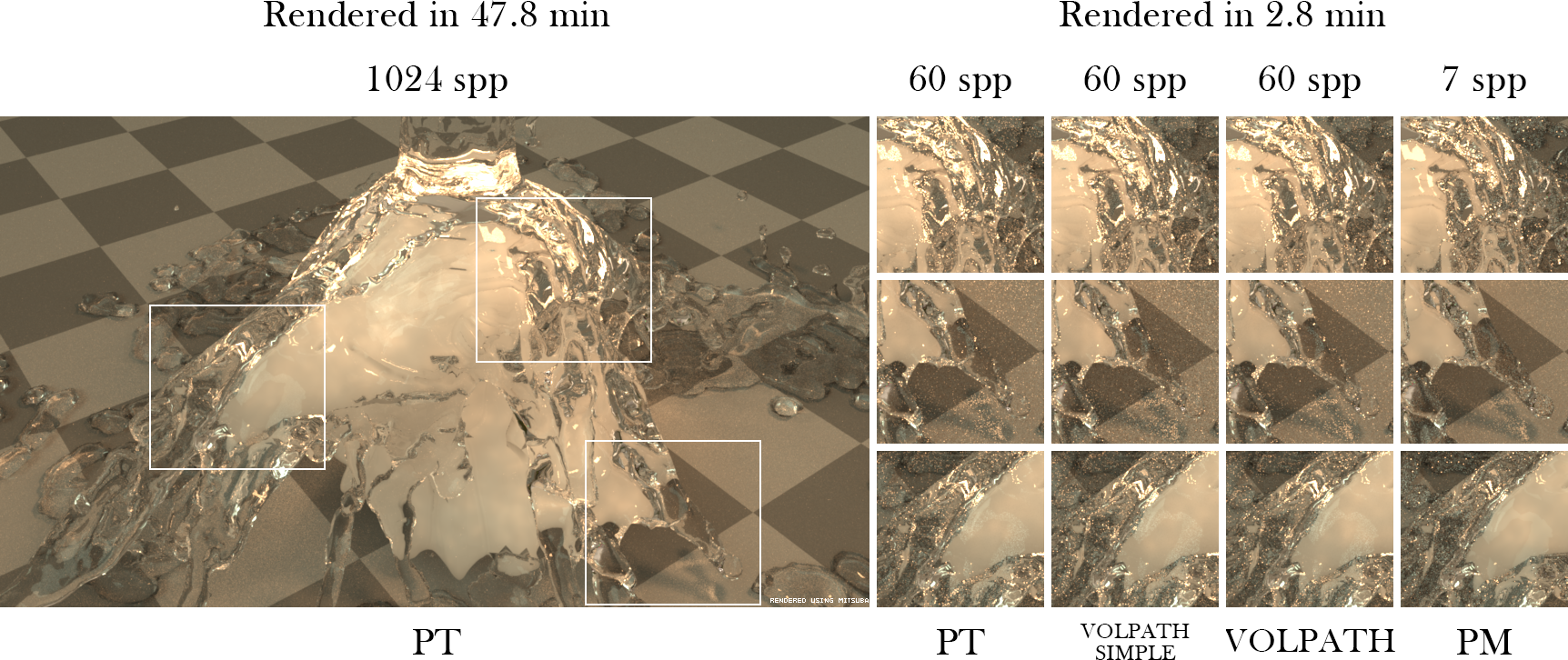}
	\caption{Reference rendered with \gls{PT} (Independent Sampler) at 1024 \gls{spp}. Comparison images (Independent Sampler) rendered in 2.8 minutes.}
	\label{fig:complex_raven_watercoat_evaluation}
\end{figure*}

The reference in Figure \ref{fig:complex_raven_watercoat_evaluation} was rendered with \gls{PT} (Independent Sampler) with 1024 samples per pixel. The comparison images were rendered in 2.8 minutes. Again the \gls{PT}, the \gls{VOLPATH} \emph{integrator}s and \gls{PM} are able to capture the \gls{SSS} materials in a scene. The results from \gls{PT}, \gls{VOLPATH} and \gls{VOLPATH_SIMPLE} had similar results. One more closeup visual inspection, the \gls{PM} seems to have achieved the best results overall. It has the lowest perceivable variance and it could resolve the \gls{SSS} material under the water better than the other \emph{integrator}s.

\subsection{Lens light transmission}
\begin{figure*}[!htbp]
	\centering
	\includegraphics[width=\linewidth]{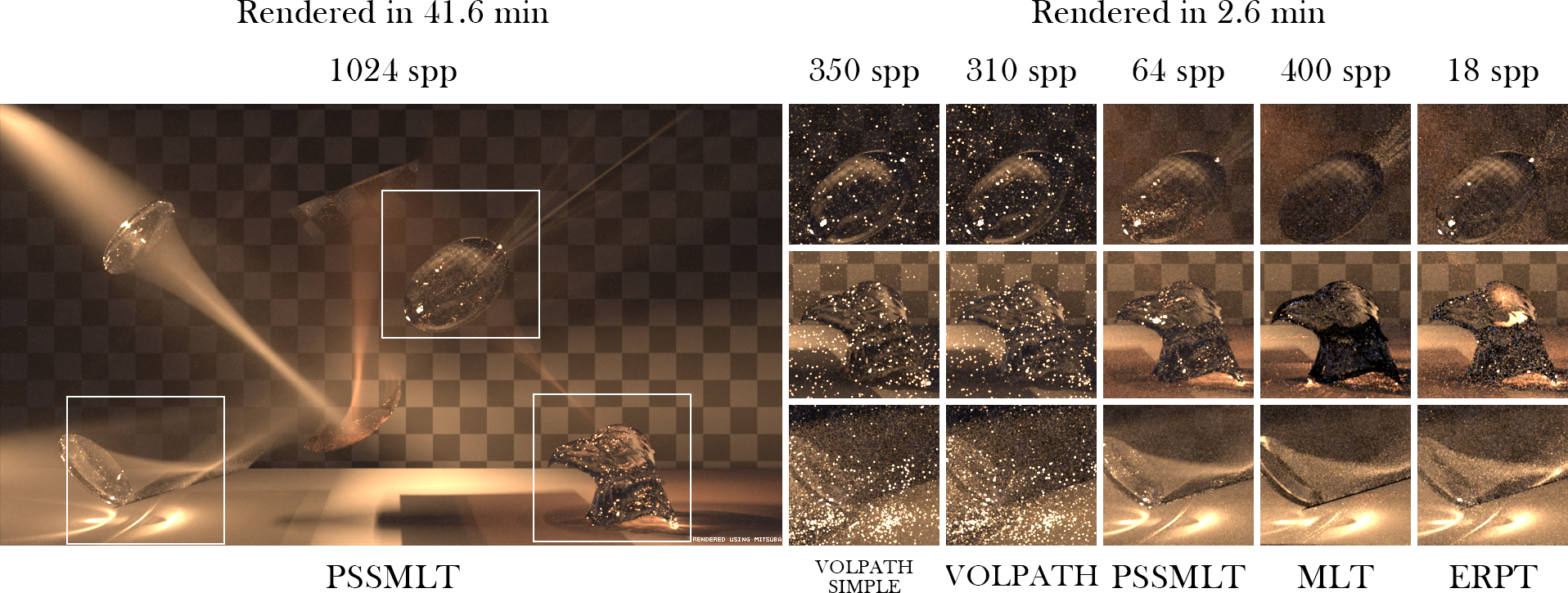}
	\caption{Reference rendered with \gls{PT} (Independent Sampler) at 1024 \gls{spp}. Comparison images (Independent Sampler) rendered in 2.6 minutes.}
	\label{fig:lenses_convex_evaluation}
\end{figure*}

The reference in Figure \ref{fig:lenses_convex_evaluation} was rendered with \gls{PSSMLT} with 1024 samples per pixel. the comparison images were rendered in 2.6 minutes. \gls{VOLPATH_SIMPLE} and \gls{VOLPATH} achieved similar results, both having high perceivable variance overall however. \gls{PSSMLT}, \gls{MLT} and \gls{ERPT} have similar results as well. The most difference can be seen for the complex glass object (Figure \ref{fig:lenses_convex_evaluation} middle row) and the thick lens (Figure \ref{fig:lenses_convex_evaluation} top row). The results from \gls{MLT} are generally too dark. The result from \gls{ERPT} has a better approximation but still has problems resolving the complex glass object. Overall, the \gls{PSSMLT} achieved the best results.

\subsection{The lens effect}
\begin{figure*}[!htbp]
	\centering
	\includegraphics[width=\linewidth]{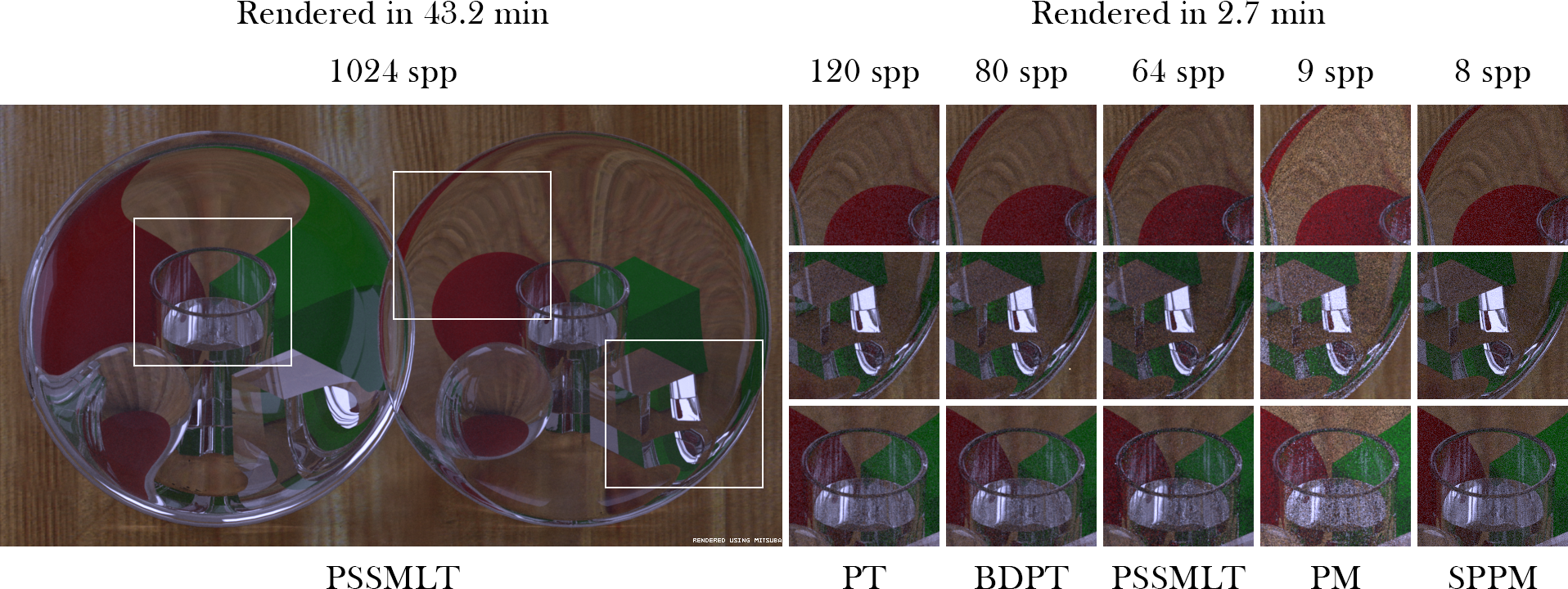}
	\caption{Reference rendered with \gls{PSSMLT} (Independent Sampler) at 1024 \gls{spp}. Comparison images (Independent Sampler) rendered in 2.7 minutes.}
	\label{fig:lenses_differentroundness_evaluation}
\end{figure*}

The reference in Figure \ref{fig:lenses_differentroundness_evaluation} was rendered with \gls{PSSMLT} with 1024 samples per pixel. The comparison images were rendered in 2.7 minutes. \gls{PT}, \gls{BDPT}, \gls{PSSMLT} and \gls{SPPM} achieved very similar results. \gls{SPPM} delivered an image that appears sharper than for the other \emph{integrator}s. The results from the \gls{PT}, \gls{BDPT} and \gls{PSSMLT} are slightly blurry. The image rendered with \gls{PM} is generally too bright and the perceivable variance is high as well. \gls{PSSMLT} had no advantage for this scene, the light energy is considerably uniform over the scene so \gls{PT} and \gls{BDPT} are preferable in terms of memory usage here. \gls{SPPM} delivers the best result but also has the highest memory usage.

\subsection{Double mirrors}
\begin{figure*}[!htbp]
	\centering
	\includegraphics[width=\linewidth]{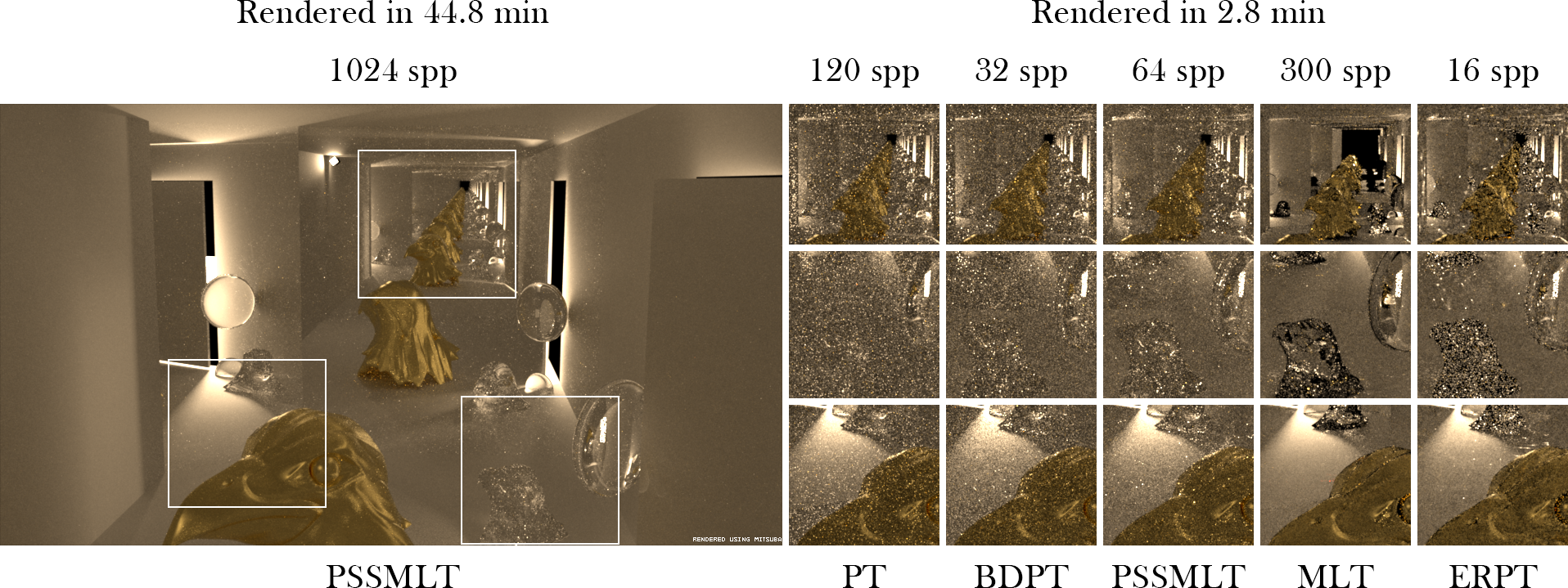}
	\caption{Reference rendered with \gls{PSSMLT} (Independent Sampler) at 1024 \gls{spp}. Comparison images (Independent Sampler) rendered in 2.8 minutes.}
	\label{fig:mirrormirror_onthewall_evaluation}
\end{figure*}

The reference in Figure \ref{fig:mirrormirror_onthewall_evaluation} was rendered with \gls{PSSMLT} with 1024 samples per pixel. The comparison images were rendered in 2.8 minutes. As expected, the \gls{PT} had troubles rendering the scene because of the relatively complex light setup. The perceivable variance is considerably high, as well as for \gls{BDPT} which has the same issues. Yet, \gls{BDPT} still exhibits slightly lower perceivable variance than \gls{PT}. Surprisingly, \gls{PSSMLT} also has troubles with the scene and also exhibits considerable variance but still has a better visual quality than \gls{PT} and \gls{BDPT}. 

The \gls{MLT} version was rendered with 300 samples per pixel and stopped when the expected render time was reached. The ground and wall portions of the scene have much less perceivable variance than for the other \emph{integrator}s. However, \gls{MLT} could not correctly resolve the interior of the complex glass object (Figure \ref{fig:mirrormirror_onthewall_evaluation} \gls{MLT} middle row). Also it could not resolve the mirror part (Figure \ref{fig:mirrormirror_onthewall_evaluation} top row) like the other \emph{integrator}s. It occurs as if the path depth is too low although it has been rendered with 32 bounces or the \gls{MLT} did not have enough time to sample these areas. \gls{ERPT} achieved a similar visual quality like \gls{PSSMLT} but with slightly less perceivable variance overall.

\subsection{Glass prisma}
\begin{figure*}[!htbp]
	\centering
	\includegraphics[width=\linewidth]{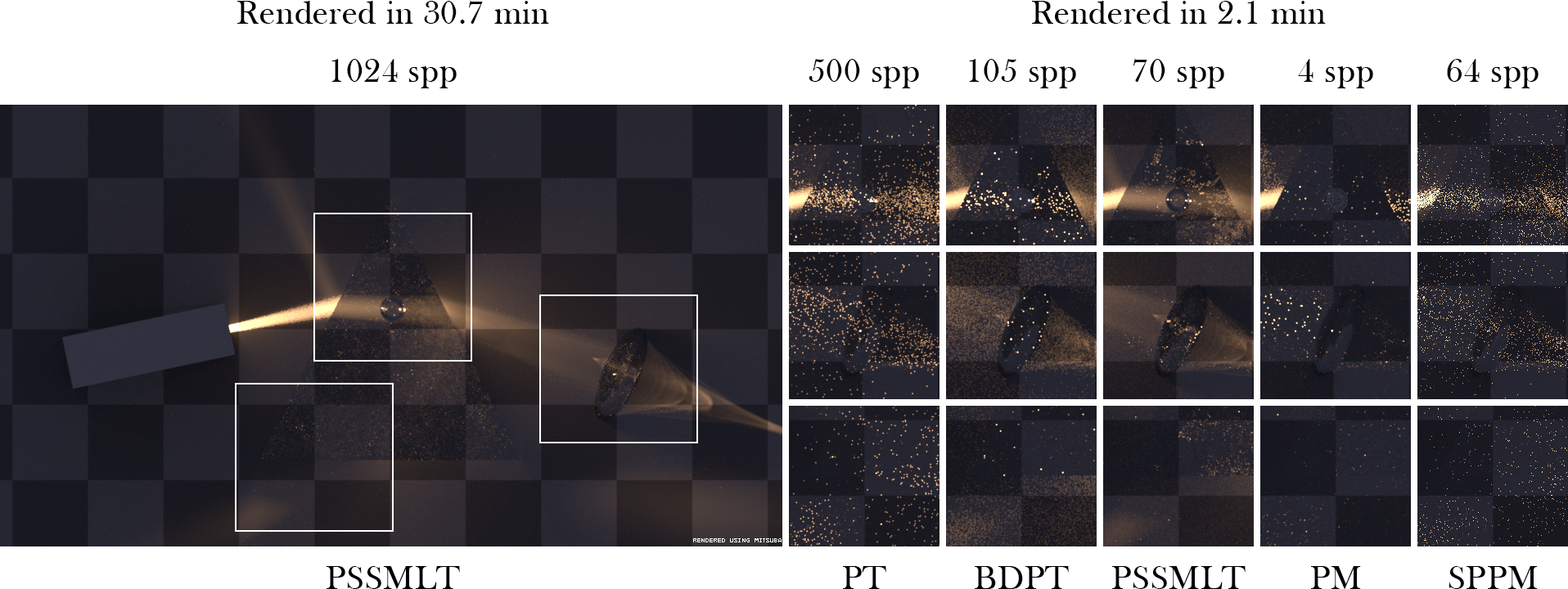}
	\caption{Reference rendered with \gls{PSSMLT} (Independent Sampler) at 1024 \gls{spp}. Comparison images (Independent Sampler) rendered in 2.1 minutes.}
	\label{fig:prisma_evaluation}
\end{figure*}

The reference in Figure \ref{fig:prisma_evaluation} was rendered with \gls{PSSMLT} with 1024 samples per pixel. The comparison images were rendered in 2.1 minutes. \gls{PT}, \gls{PM} and \gls{SPPM} had considerable problems catching light paths in this scene. Especially, in case of \gls{PM}, the scene looks too dark overall. 

The results from the \gls{BDPT} were better than for the \gls{PT}, but the perceivable variance is still high. \gls{PSSMLT} resolved the ground of the scene and the caustics from the lens relatively well (Figure \ref{fig:prisma_evaluation} \gls{PSSMLT} middle row). However, also \gls{PSSMLT} had troubles resolving the lighting within the glass materials.

\subsection{Shadows of different light sources}
\begin{figure*}[!htbp]
	\centering
	\includegraphics[width=\linewidth]{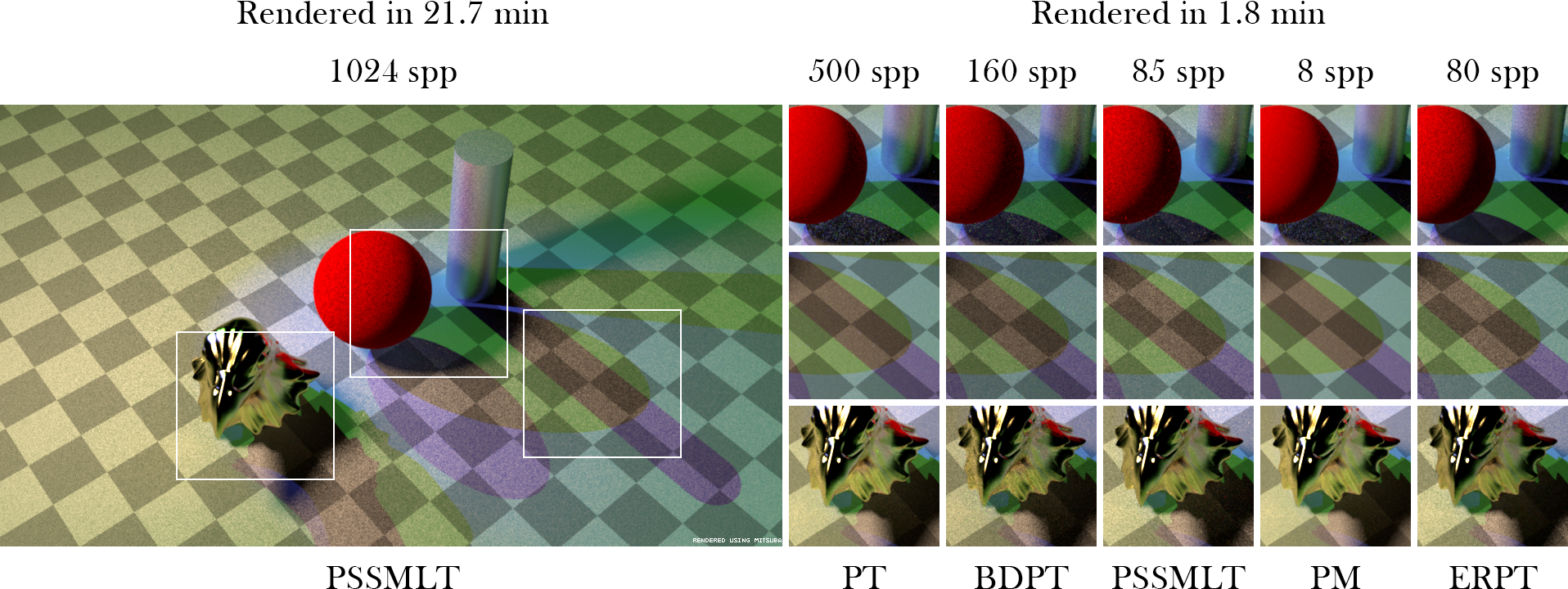}
	\caption{Reference rendered with \gls{PSSMLT} (Independent Sampler) at 1024 \gls{spp}. Comparison images (Independent Sampler) rendered in 1.8 minutes.}
	\label{fig:softshadows_evaluation}
\end{figure*}

The reference in Figure \ref{fig:softshadows_evaluation} was rendered with \gls{PSSMLT} with 1024 samples per pixel. The comparison images were rendered in 1.8 minutes. Regarding the objects, especially the red sphere and the metal cylinder (Figure \ref{fig:softshadows_evaluation} top row), the \gls{PT} and \gls{PM} resulted in images with the lowest perceivable variance.

Other than that, the results from all \emph{integrator}s are very similar. Preferable is the \gls{PM} because it has the lowest perceivable variance overall. However, the shadow of the red sphere gets resolved better with \gls{ERPT} (Figure \ref{fig:softshadows_evaluation} \gls{ERPT} top row) than for the other \emph{integrator}s.

\subsection{Light from slightly opened door}
\begin{figure*}[!htbp]
	\centering
	\includegraphics[width=\linewidth]{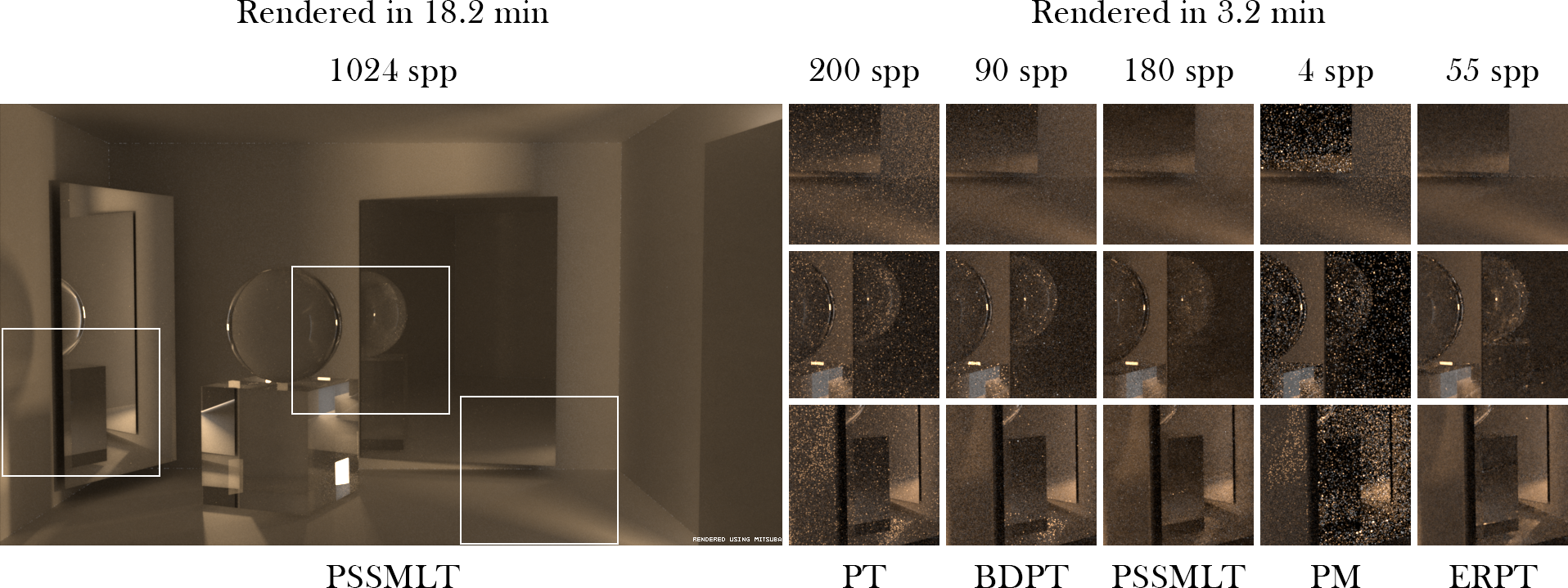}
	\caption{Reference rendered with \gls{PSSMLT} (Independent Sampler) at 1024 \gls{spp}. Comparison images (Independent Sampler) rendered in 3.2 minutes.}
	\label{fig:softshadows_lightfromdoor_evaluation}
\end{figure*}

The reference in Figure \ref{fig:softshadows_lightfromdoor_evaluation} was rendered with \gls{PSSMLT} with 1024 samples per pixel. The comparison images were rendered in 3.2 minutes. \gls{PT} again exhibits the most overall perceivable variance of all \emph{integrator}s because of the complex light setup (light coming from a door that is slightly opened). \gls{BDPT} delivers a better image than \gls{PT} in terms of variance overall. \gls{PSSMLT} is comparable to the result from \gls{BDPT} but resolves light paths a bit better (less perceivable variance) as seen in Figure \ref{fig:softshadows_lightfromdoor_evaluation} in the bottom row. 

The \gls{PM} had troubles resolving the mirror area which results in regions that are too dark and with high perceivable variance. \gls{ERPT} resolves the area in the left mirror better than \gls{PSSMLT} (Figure \ref{fig:softshadows_lightfromdoor_evaluation} bottom row). Also the perceivable variance overall is slightly lower with \gls{ERPT} than with \gls{PSSMLT}.

\subsection{Light from outside}
\begin{figure*}[!htbp]
	\centering
	\includegraphics[width=\linewidth]{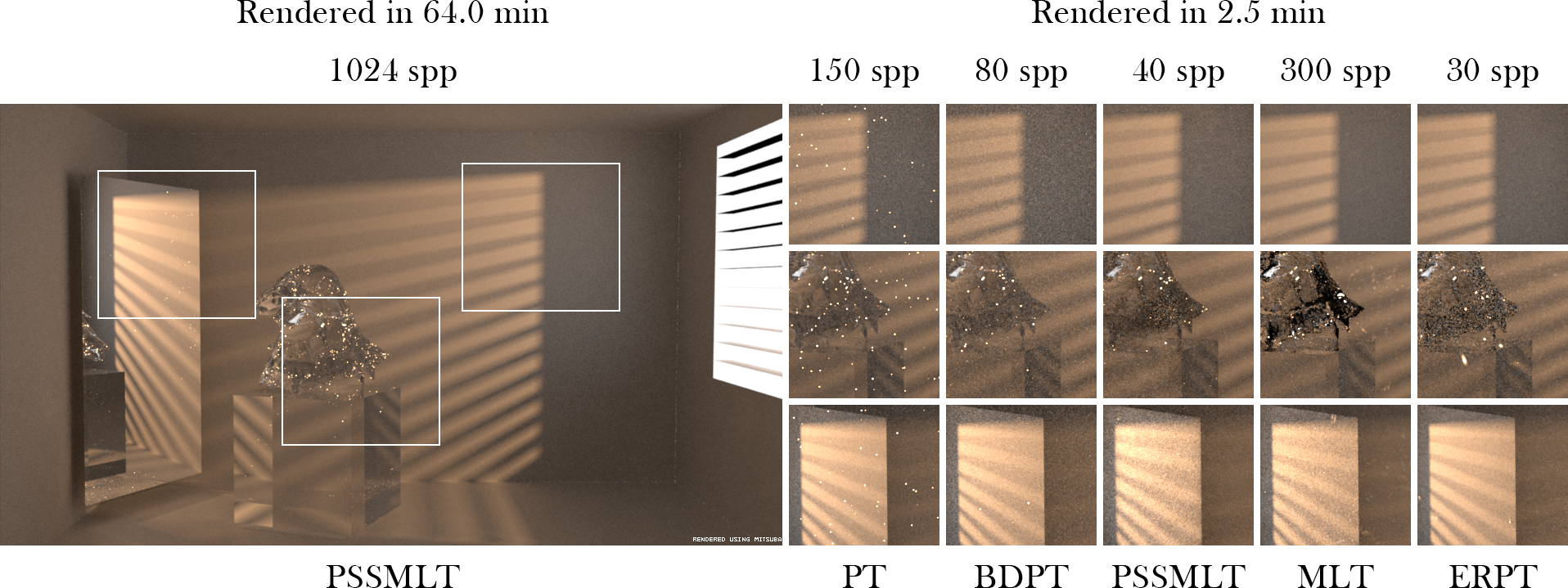}
	\caption{Reference rendered with \gls{PSSMLT} (Independent Sampler) at 1024 \gls{spp}. Comparison images (Independent Sampler) rendered in 2.5 minutes.}
	\label{fig:softshadows_lightfromoutside_evaluation}
\end{figure*}

The reference in Figure \ref{fig:softshadows_lightfromoutside_evaluation} was rendered with \gls{PSSMLT} with 1024 samples per pixel. The comparison images were rendered in 2.5 minutes. The results are similar to those of the previous scene. \gls{PT} has little perceivable variance on the wall parts of the room (Figure \ref{fig:softshadows_lightfromoutside_evaluation} \gls{PT} top and bottom row). \gls{BDPT} has less perceivable variance than \gls{PT}. As in the case of the complex glass object (Figure \ref{fig:softshadows_lightfromoutside_evaluation} middle row), it resolves the lighting in the interior slightly better than \gls{PSSMLT}. \gls{MLT} has troubles resolving the lighting of the interior of the complex glass object as well, even more than \gls{PSSMLT}. Also \gls{MLT} has a few outliers (bright blotches like in the middle row on the back wall) spread all over the scene, which are not present for the other results. \gls{ERPT} has similar results like \gls{PSSMLT} and \gls{BDPT} but exhibits higher perceivable variance for the complex glass object.

\subsection{\gls{SSS} and realistic light setup}
\begin{figure*}[!htbp]
	\centering
	\includegraphics[width=\linewidth]{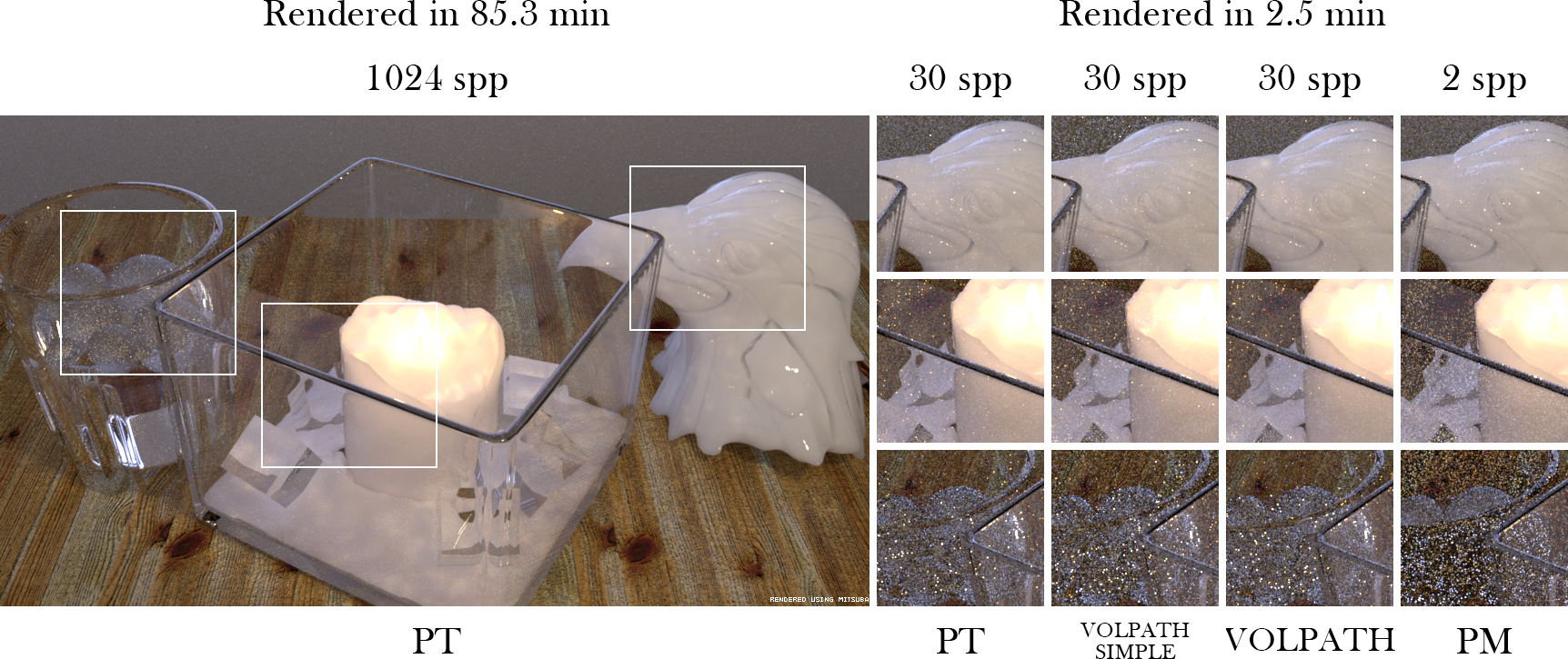}
	\caption{Reference rendered with \gls{PT} (Independent Sampler) at 1024 \gls{spp}. Comparison images (Independent Sampler) rendered in 2.5 minutes.}
	\label{fig:sss_caustics_evaluation}
\end{figure*}

The reference in Figure \ref{fig:sss_caustics_evaluation} was rendered with \gls{PSSMLT} with 1024 samples per pixel. The comparison images were rendered in 2.5 minutes. \gls{PT}, \gls{VOLPATH_SIMPLE}, \gls{VOLPATH} and \gls{PM} have very similar results. \gls{VOLPATH_SIMPLE} has slightly more perceivable variance overall than the other results. \gls{PM} has problems resolving the glass materials in the scene (Figure \ref{fig:sss_caustics_evaluation} \gls{PM} bottom row), generally they are too dark.

\subsection{Complex \gls{SSS} and rough glass surfaces}
\begin{figure*}[!htbp]
	\centering
	\includegraphics[width=\linewidth]{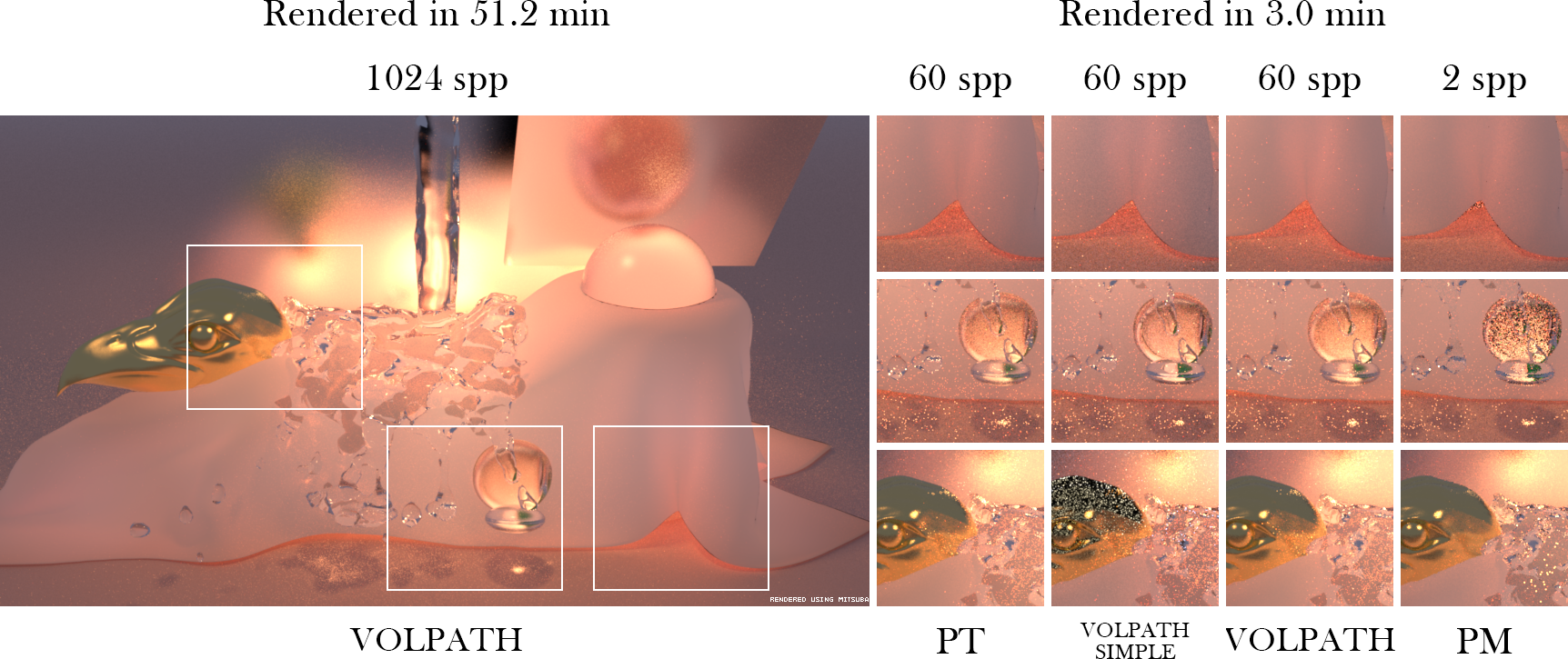}
	\caption{Reference rendered with \gls{VOLPATH} (Independent Sampler) at 1024 \gls{spp}. Comparison images (Independent Sampler) rendered in 3.0 minutes.}
	\label{fig:sss_clothcover_evaluation}
\end{figure*}

The reference in Figure \ref{fig:sss_clothcover_evaluation} was rendered with \gls{VOLPATH} with 1024 samples per pixel. The comparison images were rendered in 3.0 minutes. \gls{PT} and \gls{VOLPATH} has very similar results (because \gls{VOLPATH} uses \gls{PT} for surfaces). The perceivable variance is relatively high for the caustics on the ground coming from the water drops and lenses (Figure \ref{fig:sss_clothcover_evaluation} middle row). \gls{VOLPATH_SIMPLE} generally has very similar results like \gls{PT} and \gls{VOLPATH} as well but it has troubles rendering the top of the complex object with the gold material (Figure \ref{fig:sss_clothcover_evaluation} bottom row), it appears too dark and with high perceivable variance. \gls{PM} achieved very similar results compared to \gls{VOLPATH} and \gls{PT} as well and it has no obvious advantages or disadvantages except from the slightly higher memory usage.

\subsection{Water caustics and broad light distribution}
\begin{figure*}[!htbp]
	\centering
	\includegraphics[width=\linewidth]{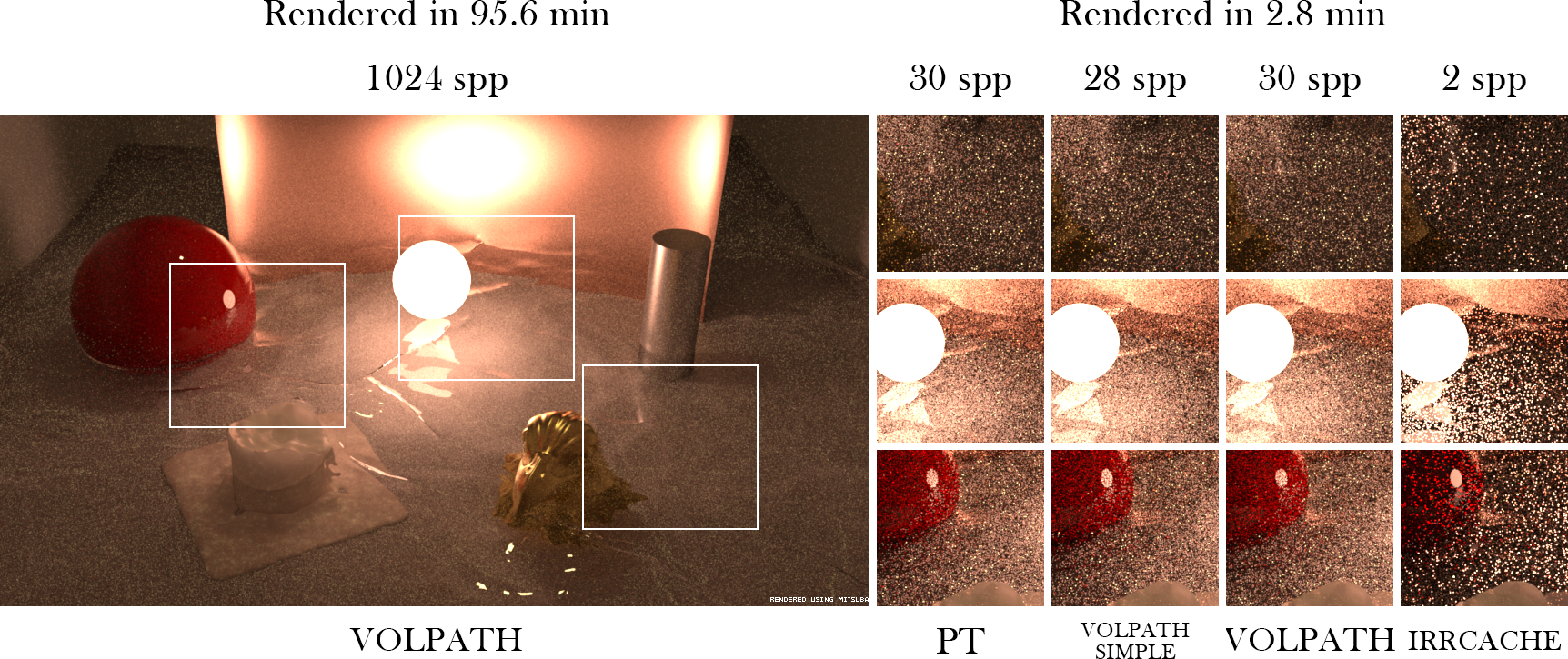}
	\caption{Reference rendered with \gls{VOLPATH} (Independent Sampler) at 1024 \gls{spp}. Comparison images (Independent Sampler) rendered in 2.8 minutes.}
	\label{fig:underwater_light_evaluation}
\end{figure*}

The reference in Figure \ref{fig:underwater_light_evaluation} was rendered with \gls{VOLPATH} with 1024 samples per pixel. The comparison images were rendered in 2.8 minutes. This scene exhibits difficult light paths so all of the \emph{integrator}s had difficulties rendering the scene. \gls{PT}, \gls{VOLPATH_SIMPLE} and \gls{VOLPATH} all show very similar results. However, the scene looks very grainy overall because of the \gls{PT} not being capable of finding relevant light paths. The problem is, that an \gls{SSS} material exists in the scene and it cannot be rendered with \emph{integrator}s like \gls{PSSMLT} which may resolve the light situation considerably better. The \gls{IRRCACHE} \emph{integrator} was used in conjunction with \gls{PM} as well but did not deliver a better result than the others. The scene looks darker overall and the perceivable variance is high as well.

\section{Conclusion}

In the previous chapter we have seen that the discussed \emph{integrator}s have different advantages and disadvantages depending on the scene setup. In conclusion, we will recapitulate the most important aspects that came to light in the evaluation chapter.

The \gls{PT} \emph{integrator} generally has problems rendering complex light conditions, this may result in high perceivable variance because the algorithm (randomized rays) has problems sampling important light paths sufficiently. Furthermore, this issue also affects rendering of glass materials and caustics -- more advanced \emph{integrator}s like \gls{PSSMLT} might be the better choice for that. Interestingly, certain light and material conditions may lead to less perceivable variance than \gls{BDPT}, as is the case in Figure \ref{fig:caustic_direct_light_waterwaves_evaluation}, mainly because \gls{PT} used far more \gls{spp} than \gls{BDPT} and most of the objects have diffuse materials which are easy to render.

In general, \gls{BDPT} has difficulties rendering translucent materials like glass as well. However, it seems to be the better choice than \gls{PT} (with exceptions like mentioned before). Under certain light and material conditions, \gls{BDPT} might come close to the visual quality of \gls{PSSMLT}, see Figure \ref{fig:softshadows_lightfromdoor_evaluation} for an example. Certain conditions can also result in \gls{BDPT} being able to resolve caustics slightly better than \gls{PSSMLT} like in Figure \ref{fig:complex_glass_evaluation}.

The \gls{PSSMLT} \emph{integrator} generally delivers good results when dealing with difficult light paths, as is the case with translucent materials in conjunction with diffuse surfaces (for example light going through the water surface, hitting the ground and going through the water again until eventually the camera catches the light).  

\gls{MLT} generally delivers similar results like \gls{PSSMLT}, however, it might also render caustics with less perceivable variance. For smooth glass materials it also generally has advantages in visual quality compared to \gls{PSSMLT}. As in the case of Figure \ref{fig:caustic_direct_light_sphere_evaluation}, \gls{MLT} may deliver results with odd behavior (\gls{MLT} bottom row) which differ considerably from the other results. Figure \ref{fig:caustic_direct_light_waterwaves_pm_evaluation} also shows that it might result in glass materials appearing darker in the rendering, this might come from the time limitation for the algorithm to converge, a good example concerning this issue can also be seen in Figure \ref{fig:mirrormirror_onthewall_evaluation}, where the glass materials are also darker and the mirror part is not completely resolved compared to the other results.

\gls{PM} often delivers visual quality that is below the results from \gls{BDPT} in the same render time. Glass materials also seem to be darker than from other \emph{integrator}s (for instance in Figure \ref{fig:caustic_direct_light_waterwaves_evaluation}), this may be because of the limited render time. \gls{PM} converges to the reference when enough photons are stored in the photon maps, ideally tending to an indefinite amount. As a disadvantage, \gls{PM} generally requires considerably more memory than other \emph{integrator}s, but it delivers good results if enough time is available to render the scene. \gls{PM} might not work well for scenes with \emph{participating media} like in Figure \ref{fig:caustic_direct_light_pendulum_pm_evaluation}.

The \gls{ERPT} \emph{integrator} is generally a good choice for scenes, especially for glass materials and complex caustics. However, it has difficulties rendering \emph{participating media} and its visual quality is below the results from \gls{PSSMLT} and \gls{BDPT}. Furthermore, it can result in visually blurry renderings, as in the case of Figure \ref{fig:colorbleeding_evaluation}. Interestingly, as in the case of Figure \ref{fig:complex_glass_evaluation}, the rough glass material appears to have the lowest perceivable variance among the tested \emph{integrator}s, but the smooth glass material appears to have the highest perceivable variance.

\gls{VOLPATH_SIMPLE} and \gls{VOLPATH} seem to have, in comparison to other \emph{integrator}s like \gls{PSSMLT}, more difficulties rendering \emph{participating media}. The difference when rendering \gls{SSS} (dipole-based subsurface scattering) materials is not noticeable compared to the results from \gls{PT} or \gls{PM} like in Figure \ref{fig:complex_sss_evaluation}, this is expected because \gls{VOLPATH} is basically a \gls{PT} with the possibility to render \emph{participating media}. \gls{SPPM} seems not to be a good choice in general, although it might deliver better results in rare cases, like in Figure \ref{fig:lenses_differentroundness_evaluation}.

The designed scenes for this paper should show where \emph{integrator}s exhibit large differences in the results. Mentionable are the scenes in Figures \ref{fig:caustic_direct_light_pendulum_pm_evaluation}, \ref{fig:caustic_direct_light_sphere_evaluation}, \ref{fig:caustic_direct_light_waterwaves_evaluation}, \ref{fig:caustic_direct_light_waterwaves_pm_evaluation}, \ref{fig:complex_glass_evaluation}, \ref{fig:mirrormirror_onthewall_evaluation} due to the large variations in the renderings. 

The scenes are all available in XML format, they can conveniently be loaded and rendered in \emph{mitsuba}. Additionally, the \emph{Blender} files are published, which contain the scene setups and a library containing the linked objects. When adding new objects to a scene, it is preferable to link the blend files instead of appending, so the material parameters can be adjusted from a single point of reference, but the objects can also be made local to test different parameters only in that scene. The scenes have been tested with \emph{Blender} 2.79b, \emph{mitsuba} version 0.5.0 and \emph{mitsuba} \emph{Blender} plugin version 0.5.1.

\section*{Acknowledgments}
Research reported in this paper was supported by Austrian Science Fund (FWF): ORD 61

\FloatBarrier

\bibliographystyle{alpha}
\bibliography{tsdpbr-paper}

\FloatBarrier
\pagebreak

\appendix

\section{Appendix}

\glsresetall 

\subsection{\glsentrylong{PBR}}

\subsubsection{\emph{Related Work}}
In this section, historic approaches and developments will be presented, concluding with recent techniques that are used in the field of \gls{PBR}. Early render systems did not have the appropriate methods and algorithms to synthesize realistic images. Rendering an image was very costly both in terms of hardware resources and algorithmic complexity. First advancements on a more realistic lighting approach has been achieved by Turner Whitted \cite{WhittedT1980} by using the idea of ray tracing to compute the distribution of light in scenes in a more realistic way. Another important step towards higher realism was made by Goral et al. \cite{GoralEtAl1984} by investigating the light exchange between surfaces in scenes. This led to the approach of \emph{radiosity} in computer graphics, which can compute the distribution of light in a scene. However, algorithms based on \emph{radiosity} were hard to handle because of the underlying computations. After several years of research work on \emph{radiosity}, \emph{path tracing} has been introduced. \cite{Kajiya86therendering}

\emph{Path tracing} simulates the distribution of light in a scene with help of \emph{Monte Carlo Integration} to approximate the fundamental \emph{light transport equation} (\gls{LTE}). Scenes could be rendered with an even more physically realistic approach than ever before. However, also \emph{path tracing} would take a large amount of computation resources because of the high complexity of the underlying algorithms, so rendering scenes at that time took long and was costly in terms of the needed hardware. In following years, more and more research work has been done to improve algorithms based on \emph{path tracing}. \cite{PharrEtAl2016}

Great efforts have been made by Eric Veach \cite{Veach1998} in terms of research on \emph{Monte Carlo Integration}. He developed \emph{bidirectional path tracing} as well as \emph{Metropolis light transport} which led to huge increases in efficiency and lower rendering times. \cite{PharrEtAl2016}

Over the years many improvements and new approaches in the field of \emph{path tracing}-based algorithms led to a variety of improvements on the integral tools. The following will cover related work on other databases and papers with \gls{PBR} background.

\subsection{Basics of \gls{PBR}}
In the following, the most important aspects of \emph{physically based rendering} systems will be discussed. Almost all of them make use of ray-tracing algorithms that simply cast rays into a scene which interact with all the objects involved \cite{PharrEtAl2016}. Since \gls{PBR} aims at realism, different combinations of materials and objects in the scene can result in complex lighting conditions that need to be calculated both fast and in high visual quality.

A great compendium for the theoretical background and practical implementations for \gls{PBR} can be found in the book \emph{Physically based rendering: From theory to implementation: Third edition}, written by Pharr et al. \cite{PharrEtAl2016}. It delivers the necessary knowledge for implementing such rendering system. Most of the following information is also extracted from this book.

\subsubsection{Light Distribution 1D example}
\begin{figure}[h]
	\centering
	\includegraphics[width=0.5\textwidth]{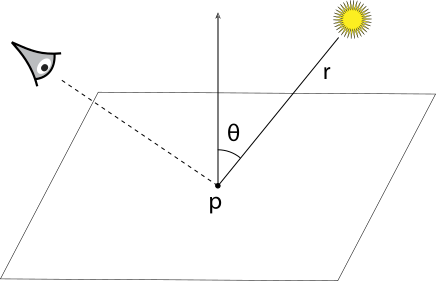}
	\caption{Geometric interpretation of light coming from a point light, where $p$ denotes the point on the surface and $r$ denotes the distance to the point light. The angle $\theta$ between $r$ and the normal at $p$ denotes the incidence angle of the light. \cite{PharrEtAl2016}}
	\label{fig:light_camera_reflection}
\end{figure}

In principle, the amount of light reflected from a point towards the camera has to be calculated. To achieve this, all the incident light at this point must be computed. Objects in real life usually emit light via different shapes, to simplify things the following will contain just a point light (that uniformly casts light rays in all directions), which does not exist in reality but can be used to observe this issue on a more abstract level. Figure \ref{fig:light_camera_reflection} shows a simple graphical interpretation of a ray from a point light hitting a surface. The intensity of the light reflection on the surface point $p$ towards the camera is what we want to calculate. \cite{PharrEtAl2016}

Because light in the real world does not perfectly reflect on surfaces but is scattered in different directions, there are several models that simulate this effect. Given the information about the location and normal of the intersection point and the location and properties of the light source, the specific material properties now define how the incident light is scattered. The so called \emph{bidirectional reflectance distribution function} (\gls{BRDF}) is often used to define these material properties. The light traveling towards the camera from that point has to be calculated by multiplying the incident light to the surface point $p$ with the corresponding \gls{BRDF} of the surface. \cite{PharrEtAl2016}

\subsubsection{Indirect Light Transport}
With the use of rays \cite{WhittedT1980}, also indirect reflections and transmissions can be simulated. Whitted's method and \emph{path tracing} solve the \emph{light transport equation} with different accuracy, Whitted ray tracing uses simplified terms for the render equation. \emph{Path tracing} solves the full equation by sending out many recursive rays per pixel (more on that later on). That means that at an intersection point another sub-ray will be cast to capture light that is incident to that point. These recursive steps make sure that the rendered image involves all the mutual reflections of the objects in the scene. However, Whitted's method only accounts for recursive evaluation of perfect refraction and reflection. \cite{PharrEtAl2016}

\subsubsection{\gls{LTE}}
As mentioned before, the \emph{light transport equation} (Equation \ref{eq:lte}) is the fundamental equation that has to be solved in order to achieve correct results in synthesized images. It models all reflections at a point on a surface, including emission coming directly from the surface and the distribution of incoming light at that point. The fact that physical realism involves taking into account all objects in the scene that radiate or transmit light, the recursive (because of sub-paths) \emph{light transport equation} has to be solved to achieve \emph{global illumination}, however this makes the numerical evaluation difficult. \cite{PharrEtAl2016}

One important principle when using the \emph{light transport equation} is \emph{energy conservation}. In terms of points on surfaces, the outgoing radiance from that point must be equal to the emitted radiance minus the absorbed energy plus the incoming light that is scattered by the surface. This is derived from the general formula of the \emph{conservation of power} (Equation 1.1 \cite{PharrEtAl2016}), where $\phi_o$ denotes the outgoing energy, $\phi_i$ the incoming energy, $\phi_e$ the emitted energy and $\phi_a$ the absorbed energy. \cite{PharrEtAl2016}

\begin{equation}
\phi_o - \phi_i = \phi_e - \phi_a
\end{equation}

Eventually, the \emph{light transport equation} for surfaces (Equation 1.2 \cite{PharrEtAl2016})  is as follows 

\begin{equation}\label{eq:lte}
L_o(p, \omega_o) = L_e(p, \omega_o) + \int_{S^2} f(p,\omega_o,\omega_i) L_i(p,\omega_i) |cos\theta_i|d\omega_i
\end{equation}

where the first term describes the emitted light and the integral describes the incoming light that is scattered on that surface point.

\subsubsection{Monte Carlo Integration solves the \gls{LTE}}
\label{section:monte_carlo_integration}
The following will give a brief overview on \emph{Monte Carlo Integration}, which is fundamental for \emph{path tracing}, since the underlying \emph{light transport equation} cannot be computed analytically but only numerically using this method.

\emph{Monte Carlo Integration} involves computations that approximate the integral of a given function by averaging randomly chosen samples. In the context of \emph{path tracing}, this idea is applied by letting the algorithm go over one pixel several times, each time computing a random sample (light path). Statistically, averaging the different values from different runs will converge to the correct value. \cite{PharrEtAl2016}

In detail, \emph{Monte Carlo Integration} aims at solving an arbitrary integral, in terms of 1D let it be $\int_{a}^{b} f(x) dx$. The expected value of the \emph{Monte Carlo Estimator} $E[F_n]$ (Equation 1.3 \cite{PharrEtAl2016}) estimates this integral.

\begin{equation}
E[F_N] = E\bigg[\frac{b-a}{N} \sum_{i=1}^{N} f(X_i)\bigg]
\end{equation}

where $X_i \in [a,b]$ and the probability is uniformly distributed and $N$ denotes the number of random samples. In order to reduce variance from the integration, the underlying probability density function $p(x)$ can be chosen heuristically to address this issue (\emph{Importance Sampling}). This leads to the adapted estimator (Equation 1.4 \cite{PharrEtAl2016}).

\begin{equation} \label{eq:mc_estimator}
E[F_N] = E\bigg[\frac{1}{N} \sum_{i=1}^{N} \frac{f(X_i)}{p(X_i)}\bigg]
\end{equation}

The estimator can easily be used with higher dimensions, making it the only practical numerical method that converges high dimensional integrals independently from the dimensionality, that means the size of the dimension does not have impact on the performance of \emph{Monte Carlo Integration}. 

One negative aspect is that the error behaves inversely proportional to the amount of samples in a way that if the error should be halved, four times as many samples have to be evaluated. Graphically, approximation errors from the integration result in noisy pixels that are either too dark or too bright. \cite{PharrEtAl2016}

\subsubsection{Advanced sampling methods}

\emph{Path tracing} was the first algorithm that solves the full \emph{light transport equation} numerically. However, it can lead to partially very grainy images due to high variance in more complex lighting conditions \cite{PharrEtAl2016}. The following will give a brief overview on more advanced sampling techniques that are generally better than pure \emph{path tracing} in terms of render quality and convergence time.

\paragraph{\gls{NEE}}

\emph{Next Event Estimation} is a sampling technique that aims at reducing variance of \emph{Monte Carlo}-sampling. Basically, \gls{NEE} achieves this by searching for direct light sources at any path intersection. This can improve render performance and quality considerably when applied at every surface intersection of the path. \cite{Koerner16}

\paragraph{Importance Sampling} 
\label{subsubsection:importance_sampling}
\emph{Importance Sampling} is a method for \emph{Monte Carlo Integration} to reduce variance and convergence as well. The \emph{Monte Carlo Estimator} from Equation \ref{eq:mc_estimator} has the property that it converges faster if the samples are taken from a \emph{probability density function} that is similar to $f(x)$ in the integrand. Graphically, this means, the focus is on areas with high (relevant) values and therefore, calculations are faster and variance of the image decreases. \cite{PharrEtAl2016}

\paragraph{\gls{BDPT}}
\label{subsubsection:bidirectional_sampling}
\gls{BDPT} is an extension to the common \emph{path tracing} algorithm. It was developed independently by Lafortune and Willems \cite{Lafortune98} and Veach and Guibas \cite{Veach95}. It is called bidirectional because it does not only cast rays going out from the camera but also from the light sources. Basically, the algorithm knows all intersection points and attempts to connect pairs of them for each camera/light path. It then evaluates if a connection between pairs of intersections is not interrupted by another object. If this is the case, the corresponding path is added to the light estimation. \gls{BDPT} generally speeds up the convergence of the light computation while also delivering less variance with fewer samples per pixel than \emph{path tracing}. \cite{PharrEtAl2016}

\begin{figure}[h]
	\centering
	\includegraphics[width=1.0\linewidth]{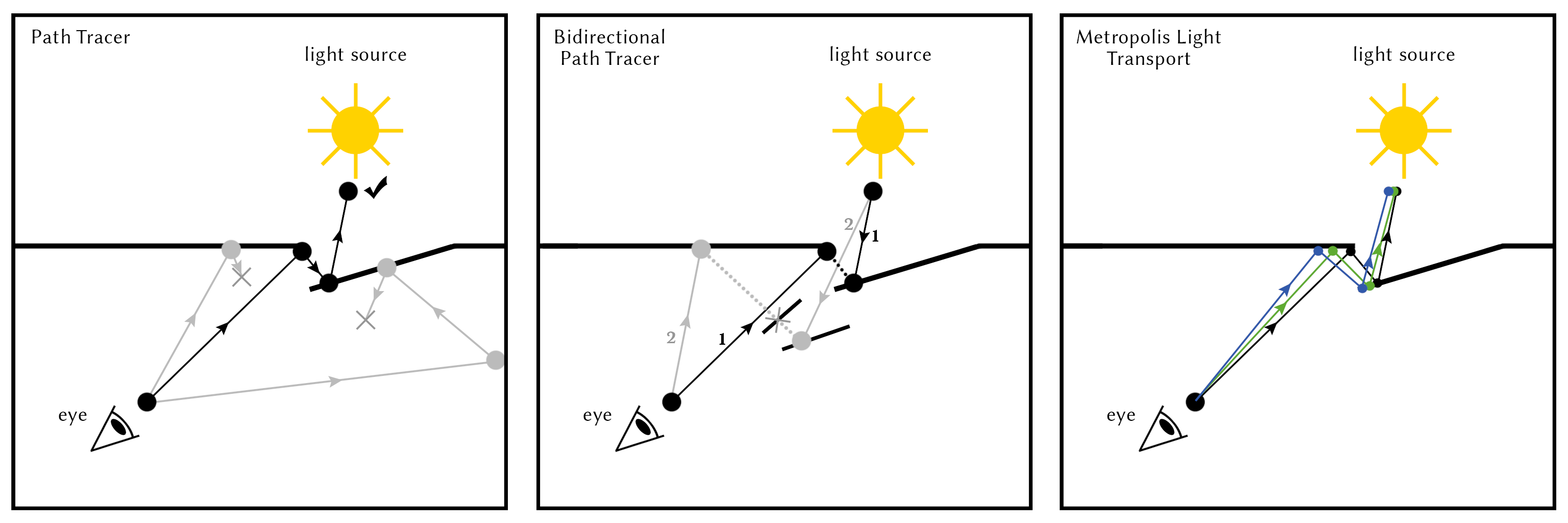}
	\caption{Graphical interpretations of light sampling via \emph{Path Tracer}, \emph{Bidirectional Path Tracer} and \emph{Metropolis Light Transport}. In the case of the \emph{Path Tracer}, the black path shows a successful capture of the light source. The gray paths depict non-successful samples, which is common for complex light conditions. For \gls{BDPT}, a ray is shot from the eye (camera) and from the light source, it then tries to connect a pair of path nodes. The black path (1) shows successfully captured light. The gray path (2) shows a failed connection because an obstacle is in-between. For \gls{MLT}, mutated paths are constructed, based on the initial black path, to search in the neighbored regions.}
	\label{fig:lightsampling} %
\end{figure}

\gls{BDPT} has the advantage that the search for a relevant light source is easier unlike for \emph{Path tracing}. A standard \emph{Path tracer} shoots random rays into the scene trying to find light sources. Because it is very rare to find light sources with random rays in scenes with difficult light setups (for instance a light that is partially covered), resulting images often exhibit high variance. \gls{BDPT} simplifies this process and generally delivers images with less variance. The basic differences of light sampling between \emph{Path tracing} and \gls{BDPT} can be seen in Figure \ref{fig:lightsampling}.

\paragraph{\gls{MLT}}

\gls{MLT} was first proposed by Veach and Guibas \cite{Veach97} in 1997. In contrast to the other methods, \gls{MLT} is not based on the \emph{Monte Carlo} method, specifically it creates samples that are statistically correlated. Basically, \gls{MLT} sequentially shoots rays into the scene. Each subsequent ray is a mutation of the previous one using \emph{Markov chain} techniques (the next sample state is dependent on the previous one). The main advantage of this method is that if a light path with high relevance is found, the following rays will search for further relevant paths in the neighborhood. The amount of searches in a region is therefore dependent on the relevance of the particular region in the scene. The basic principle of this method can be seen in Figure \ref{fig:lightsampling}. This means that \gls{MLT} is considerably good for scenes with difficult light situations like caustics where many light rays are located in small areas. However, \gls{MLT} does have performance deficiencies when it comes to relatively simple and balanced light conditions. \cite{PharrEtAl2016}

There are several other sampling strategies that have different advantages and disadvantages, however this will not be covered in detail. Especially, \gls{BDPT} and \gls{MLT} will be interesting related to the scene data set in the next chapter, because the light setups often require appropriate samplers to achieve good results in a reasonable time. Specifically, there will be scenes that have complex materials and light setups which can be easier or more difficult to sample depending on the chosen \emph{integrator}. 

The reason for this is illustrated by a simplified example showing a difficult scene setup in Figure \ref{fig:lightsampling}. It shows the comparisons of the sampling strategies that were mentioned previously. The scenes described in this paper aim at challenging these strategies in order to understand their differences and examine potential shortcomings and problems. Difficulties in sampling will later be seen as variance in the renderings, depending on the sampling technique, the algorithms may give considerably different results.

\end{document}